\documentclass[conference]{IEEEtran}
\IEEEoverridecommandlockouts
\usepackage{hyperref}
\usepackage{graphicx}
\usepackage{booktabs}
\usepackage{subcaption}
\usepackage{tabularx}
\usepackage{multirow}
\usepackage{tcolorbox}
\usepackage{enumitem}
\usepackage{array}
\usepackage{url}
\usepackage{multicol}
\usepackage{caption}
\usepackage{geometry}
\usepackage[normalem]{ulem}
\usepackage{mdframed}
\usepackage{soul}
\usepackage{url}
\usepackage{cite}
\usepackage{amsmath,amssymb,amsfonts}
\usepackage{algorithmic}
\usepackage{graphicx}
\usepackage{textcomp}
\usepackage{xcolor}
\def\BibTeX{{\rm B\kern-.05em{\sc i\kern-.025em b}\kern-.08em
    T\kern-.1667em\lower.7ex\hbox{E}\kern-.125emX}}

\definecolor{lightgray}{RGB}{230, 230, 230}
\definecolor{lightergray}{RGB}{235, 235, 235}
\definecolor{darkgray}{RGB}{150, 150, 150}
\definecolor{highlightcolor}{RGB}{0, 150, 200}

\newcommand{\code}[1]{\sethlcolor{lightergray}\texttt{\small\hl{#1}}}

\newtcolorbox{highlighted}[1]{%
    colback=lightgray,
    colframe=darkgray,
    boxrule=0.5pt,
    arc=2.5pt,
    boxsep=2pt,
    left=3pt,
    right=3pt,
    top=2pt,
    bottom=2pt,
    #1
}

\begin{document}

\title{SECURE: Benchmarking Large Language Models for Cybersecurity \thanks{ The work presented in this paper was partially supported by the U.S. Department of Energy, Office of Science under DOE contract number DE-AC02-06CH11357. The submitted manuscript has been created by UChicago Argonne, LLC, operator of Argonne National Laboratory. Argonne, a DOE Office of Science laboratory, is operated under Contract No. DE-AC02-06CH11357. The U.S. Government retains for itself, and others acting on its behalf, a paid-up nonexclusive, irrevocable worldwide license in said article to reproduce, prepare derivative works, distribute copies to the public, and perform publicly and display publicly, by or on behalf of the Government. }}

\author{\IEEEauthorblockN{1\textsuperscript{st} Dipkamal Bhusal\IEEEauthorrefmark{1}\thanks{\IEEEauthorrefmark{1} Equal contribution.}}
\IEEEauthorblockA{\textit{Rochester Institute of Technology} \\
Rochester, USA \\
db1702@rit.edu}
\and
\IEEEauthorblockN{2\textsuperscript{nd} Md Tanvirul Alam\IEEEauthorrefmark{1}}
\IEEEauthorblockA{\textit{Rochester Institute of Technology} \\
Rochester, USA \\
ma8235@rit.edu}
\and
\IEEEauthorblockN{3\textsuperscript{rd} Le Nguyen}
\IEEEauthorblockA{\textit{Rochester Institute of Technology} \\
Rochester, USA \\
ln8378@rit.edu}
\and
\IEEEauthorblockN{4\textsuperscript{th} Ashim Mahara}
\IEEEauthorblockA{\textit{RIT} \\
Rochester, USA \\
am7539@g.rit.edu}
\and
\IEEEauthorblockN{5\textsuperscript{th}  Zachary Lightcap}
\IEEEauthorblockA{\textit{RIT} \\
Rochester, USA \\
ztl1776@rit.edu}
\and
\IEEEauthorblockN{6\textsuperscript{th} Rodney Frazier}
\IEEEauthorblockA{\textit{RIT} \\
Rochester, USA \\
rlf9328@rit.edu}
\and
\IEEEauthorblockN{7\textsuperscript{th} Romy Fieblinger}
\IEEEauthorblockA{\textit{RIT} \\
Rochester, USA \\
rf7344@rit.edu}
\and
\IEEEauthorblockN{8\textsuperscript{th} Grace Long Torales}
\IEEEauthorblockA{\textit{RIT} \\
Rochester, USA \\
gtl1500@rit.edu}
\and
\IEEEauthorblockN{9\textsuperscript{th} Benjamin A. Blakely}
\IEEEauthorblockA{\textit{Argonne National Lab} \\
Lemont, USA \\
bblakely@anl.gov}
\and
\IEEEauthorblockN{ 10\textsuperscript{th} Nidhi Rastogi}
\IEEEauthorblockA{\textit{Rochester Institute of Technology (RIT)} \\
Rochester, USA \\
nxrvse@rit.edu}
}

\maketitle

\begin{abstract}
Large Language Models (LLMs) have demonstrated potential in cybersecurity applications but have also caused lower confidence due to problems like hallucinations and a lack of truthfulness. Existing benchmarks provide general evaluations but do not sufficiently address the practical and applied aspects of LLM performance in cybersecurity-specific tasks. To address this gap, we introduce the SECURE (Security Extraction, Understanding \& Reasoning Evaluation), a benchmark designed to assess LLMs performance in realistic cybersecurity scenarios. SECURE includes six datasets focused on the Industrial Control System sector to evaluate knowledge extraction, understanding, and reasoning based on industry-standard sources. Our study evaluates seven state-of-the-art models on these tasks, providing insights into their strengths and weaknesses in cybersecurity contexts. We also offer recommendations for improving LLMs reliability as cyber advisory tools and release our benchmark datasets and framework for community use at \textcolor{blue}{\url{https://github.com/aiforsec/SECURE}}. 
\end{abstract}

\begin{IEEEkeywords}
Large Language Models, Cybersecurity, Benchmarking, Dataset, Industrial Control System.
\end{IEEEkeywords}

\section{Introduction}
Recent breakthroughs in large language models (LLM) like OpenAI's ChatGPT~\cite{gpt4} have opened up their applications in many domains, including security \cite{zaboli2023chatgpt}. These models, trained on vast datasets, can generate, understand, and reason across a multitude of domains \cite{bubeck2023sparks}. Users can interact with these models in a conversational style through a simple web interface and obtain answers to their questions in a short time. Despite demonstrating high potential, LLMs are plagued by issues such as hallucinations \cite{weidinger2021ethical} and truthfulness \cite{madaan2022text}. Hence, a standard evaluation benchmark is crucial to evaluate the reliability of such models. Several benchmarks like GLUE \cite{wang2018glue}, MMLU \cite{hendrycks2020measuring}, Helm \cite{liang2022holistic} and KOLA \cite{yu2023kola}, provide standard datasets and tasks to evaluate general-purpose understanding and capabilities of LLMs. GLUE assesses LLMs' performance in understanding language, while MMLU and HELM offer a holistic evaluation across various domains like mathematics, history, computer science, and law. KOLA focuses on tasks designed to measure the cognitive abilities of LLMs.

\begin{figure}[]
    \centering
        \includegraphics[width=0.36\textwidth]{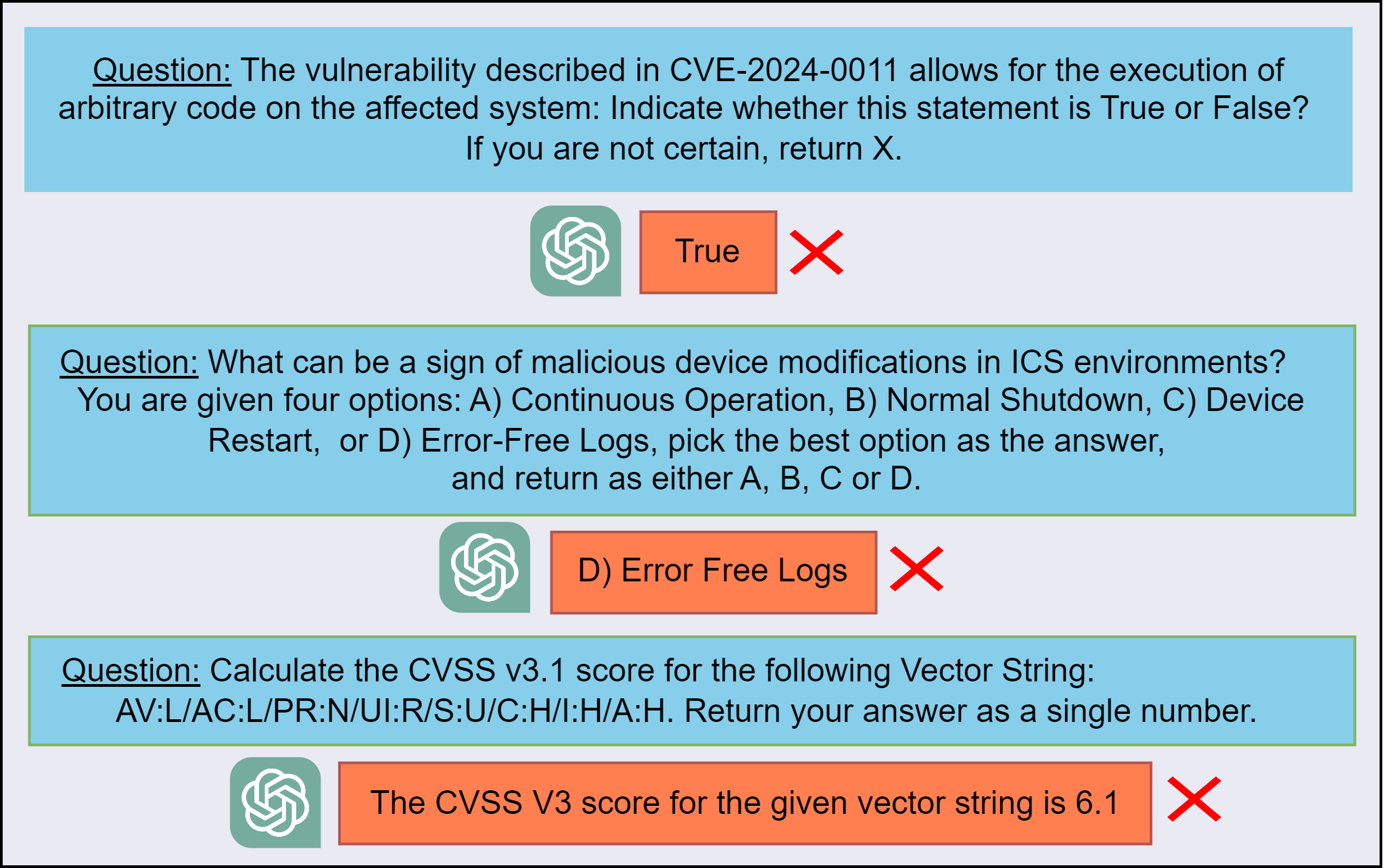}
        \caption{Conversation between a user and a ChatGPT-3.5 on ICS security questions.}
        \label{fig:motivation}
\end{figure}

Despite these advancements, there remains a significant gap in the evaluation of LLMs specifically tailored for security industries such as information security, network security, and critical infrastructure protection. Traditional benchmarks often fail to capture the practical and applied aspects of cybersecurity, leading to an incomplete assessment of LLM capabilities. These benchmarks typically focus on general language tasks and do not address specific challenges such as recognizing emerging threats, handling specialized terminology, or performing tasks like vulnerability assessment and incident response. More practical, domain-specific, and comprehensive evaluations are necessary to understand LLM performance in realistic cybersecurity scenarios \cite{sei2024considerations}. In Figure \ref{fig:motivation}, we demonstrate a conversation between ChatGPT-3.5 and a user based on our benchmark dataset (see Section \ref{sec:benchmark}). ChatGPT-3.5 model was prompted to act as a security expert, and yet, all three responses, spanning different kinds of tasks, were incorrect, showing the unreliability of these models in cybersecurity. Especially, on the first task, we ask questions about vulnerabilities discovered in 2024. Even though the model is not trained on any such recent data, it confidently makes a decision, despite our instructions to return 'X' when it is not confident.
\begin{figure}[]
    \centering
        \includegraphics[width=0.5\textwidth]{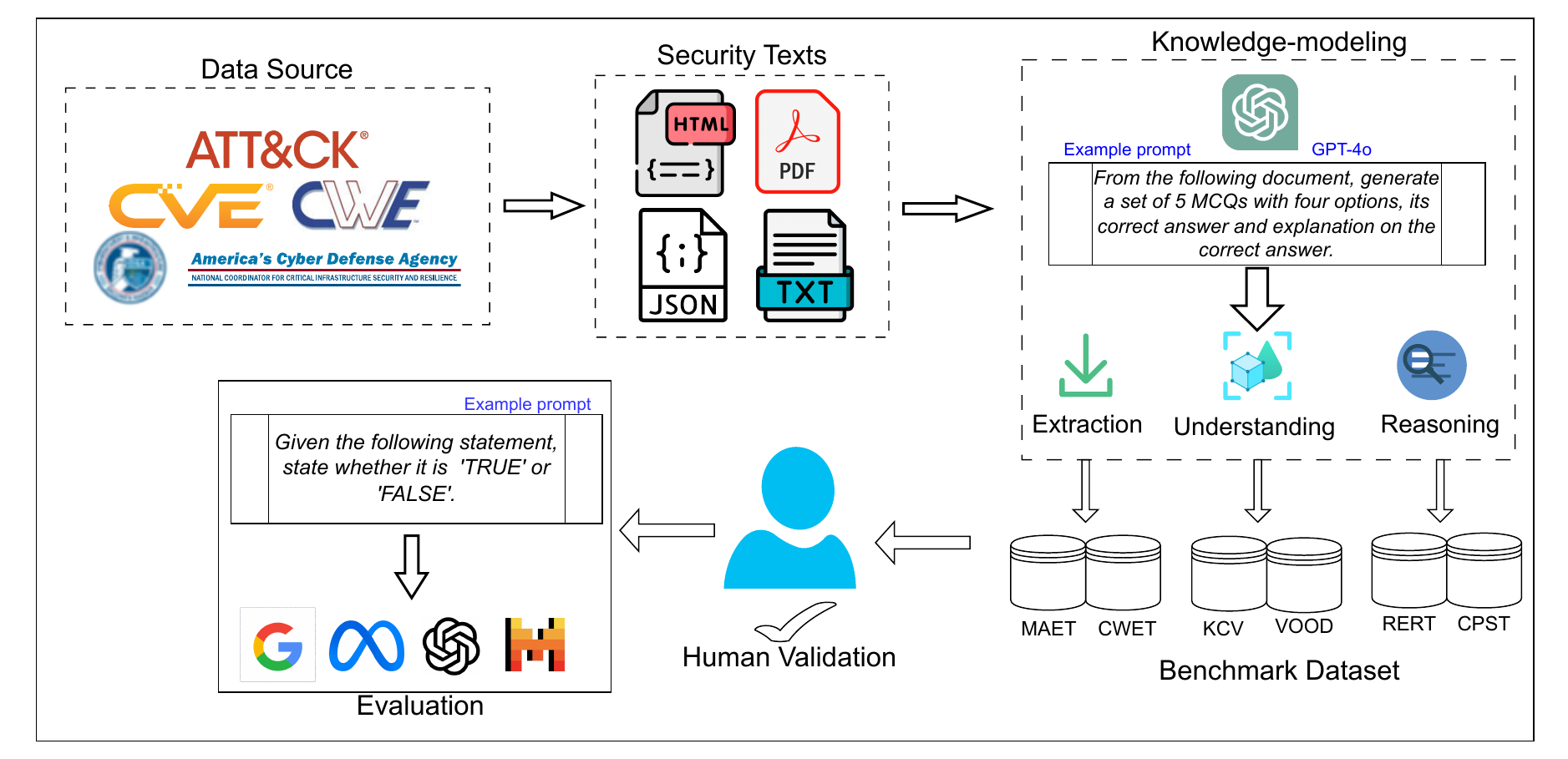}
        \caption{Overview of the SECURE benchmark.}
        \label{fig:securebenchmark}
\end{figure}
        
To address this gap, we introduce a comprehensive benchmarking framework encompassing real-world cybersecurity scenarios, practical tasks, and applied knowledge assessments. We introduce \underline{\textbf{S}}ecurity \underline{\textbf{E}}xtra\underline{\textbf{C}}tion, \underline{\textbf{U}}nderstanding \& \underline{\textbf{R}}easoning \underline{\textbf{E}}valuation (SECURE) for large language models. This benchmark provides a holistic evaluation of LLMs performance in cybersecurity applications, ensuring they meet the high standards required for deployment in critical infrastructure environments, such as Industrial Control Systems (ICS). Figure \ref{fig:securebenchmark} shows an overview of our proposed benchmark. SECURE incorporates knowledge modeling to design six different datasets (MAET: Mitre Attack Extraction Task, CWET: Common Weakness Extraction Task, KCV: Knowledge test on Common Vulnerabilities, VOOD: Vulnerability Out-of-Distribution task, RERT: Risk Evaluation Reasoning Task, and CPST: CVSS Problem Solving Task) focused on three types of knowledge evaluations: extraction, understanding, and reasoning. These datasets are sourced from standard sources, such as MITRE \cite{mitre-techniques} \cite{mitre-mitigations}, CVE \cite{cve2024}, CWE \cite{cwe1358}, and Cybersecurity and Infrastructure Security Agency (CISA) \cite{cisa_advisories}. 

\subsection*{Main Contributions}
\begin{enumerate}[leftmargin=*, noitemsep, topsep=0pt, partopsep=0pt]
\item \textbf{Assessment of Cybersecurity Knowledge:} We evaluate the cybersecurity knowledge of LLMs in terms of their capacity to assist security analysts. Specifically, we design a set of benchmark tasks tailored for the evaluation of LLMs within the context of Industrial Control Systems (ICS) cyber advisories (see Section \ref{sec:benchmark}). These tasks are grounded in comprehensive protocol specifications, which include communication standards, Common Vulnerabilities and Exposures (CVEs) with contextual details on exploitability and severity, and remediation strategies, such as patch notes and mitigation steps informed by the MITRE framework.

\item \textbf{Evaluation of LLMs}: Recent works have shown that customized LLMs introduce a loss of safety measures \cite{qi2023fine}, \& increased hallucinations \cite{gekhman2024does}. In addition, the computational and technical requirements of building customized LLMs can push researchers and industries to use commercial LLMs. However, before we can confidently claim the application of general-purpose LLMs in a specific domain, we first need to evaluate it on domain-specific tasks. To this end, we assess the performance of seven state-of-the-art, open-source, and proprietary LLMs on our cybersecurity benchmark tasks (see Section \ref{sec:experiments}). These models include ChatGPT-4 \cite{gpt4}, ChatGPT-3.5 \cite{gpt3.5}, Llama3-70B \cite{llama370b}, Llama3-8B \cite{llama38b}, Gemini-Pro \cite{gemini}, Mistral-7B \cite{jiang2023mistral}, and Mixtral-8x7B \cite{mixtral8x7b}. Following this, we customize Llama3-8B using the Retrieval-Augmented Generation (RAG) framework and fine-tuning and re-evaluate its performance on the benchmark (see Section \ref{sec:customizellm}).

\item \textbf{Insights and Recommendations:} Our evaluation reveals that, although LLMs exhibit some competence in cybersecurity tasks, their application as advisory tools demands careful consideration. We present key insights and offer recommendations to improve their practical usability (see Section \ref{sec:resultsAnalysis}).

\item \textbf{Benchmark Dataset and Framework:} We present a comprehensive set of benchmark datasets—MAET (Mitre Attack Extraction Task), CWET (Common Weakness Extraction Task), KCV (Knowledge Test on Common Vulnerabilities), VOOD (Vulnerability Out-of-Distribution), RERT (Risk Evaluation Reasoning Task), and CPST (CVSS Problem Solving Task)—along with an evaluation framework. These resources are made available to the security community to facilitate the open evaluation of future LLMs across various security-related tasks.
\end{enumerate}


\section{Background and Related Work}\label{sec:related}

\textbf{Large Language Model (LLM) in Security:} LLMs, trained on vast amounts of textual data, are capable of producing coherent and contextually relevant text. While earlier BERT-language models like SecureBERT \cite{aghaei2022securebert} and CySecBERT \cite{bayer2024cysecbert} were used for language modeling in cybersecurity, the release of GPT (Generative Pre-trained Transformer) models has changed the nature of language models. There are two types of LLMs in the industry. Open Source LLMs like Llama \cite{llama370b} and Mixtral \cite{jiang2023mistral} make their models public so they can be fine-tuned for specific downstream tasks. Closed-source LLMs like ChatGPT \cite{gpt4} and Gemini \cite{team2024gemma} allow restricted access through APIs. The most notable language models in security are code-based LLMs such as  CodeLlama \cite{roziere2023code}, adapted to analyze and generate secure code. Open-sourced models have also been fine-tuned for several cybersecurity tasks like vulnerability detection \cite{shestov2024finetuning, ferrag2023securefalcon, fang2024llm, li2024llm}, program repair \cite{silva2023repairllama}, IT operations \cite{guo2023owl}, and security knowledge assistance \cite{sultana2023towards}.

\textbf{Evaluation Benchmark:} While there are various benchmarks like GLUE \cite{wang2018glue}, MMLU \cite{hendrycks2020measuring}, Helm \cite{liang2022holistic} and KOLA \cite{yu2023kola}  for evaluating general-purpose LLMs, comprehensive cybersecurity-specific benchmarks remain limited. Existing approaches to evaluating LLMs in security tend to focus on factual knowledge rather than applied, practical cybersecurity tasks. For example: the CyberMetric dataset \cite{tihanyi2024cybermetric} comprises 10,000 questions sourced from cybersecurity standards. CyberBench \cite{liucyberbench} comprises ten datasets from different tasks namely, named-entity recognition, summarization, multiple choice, and classification. SecEval \cite{li2023seceval} evaluates cybersecurity knowledge in LLMs with 2000 multiple-choice questions across Software Security, Application Security, System Security, Web Security, Cryptography, Memory Safety, Network Security, and PenTest. SecQA \cite{liu2023secqa} consists of multiple-choice questions based on the "Computer Systems Security: Planning for Success" textbook to evaluate LLM's understanding of security principles. NetEval \cite{miao2023empirical} evaluates the knowledge of large language models in IT operational tasks within a multilingual context. OpsEval \cite{liu2023opseval} also contains multi-choice questions designed for fault root cause analysis, operational script generation, and alert information summarization. Ullah et al. \cite{ullah2024llms} tests LLMs on identifying and reasoning about software vulnerabilities using 228 code scenarios. Their findings suggest existing LLMs are unreliable in identifying vulnerabilities in source code. CTIBench \cite{alam2024ctibench} is a benchmark of four tasks to evaluate the ability of LLMs in cyber threat intelligence (CTI) landscape.

\section{Proposed Benchmark: SECURE}\label{sec:benchmark}
Consider an LLM as a cybersecurity advisor in an organization facing diverse threats, from malware to advanced persistent threats. To mitigate these risks, cybersecurity professionals must remain constantly informed about threat intelligence, security best practices, and incident response strategies specific to their organization. LLMs have the potential to leverage their vast knowledge base and natural language processing capabilities and assist security teams in identifying vulnerabilities, interpreting threat reports, and suggesting proactive measures to fortify an organization's defenses. 

However, the truthfulness and reliability of the information provided by the LLMs acting as cybersecurity advisors is crucial. To achieve this, we propose \underline{\textbf{S}}ecurity \underline{\textbf{E}}xtra\underline{\textbf{C}}tion, \underline{\textbf{U}}nderstanding \& \underline{\textbf{R}}easoning \underline{\textbf{E}}valuation (SECURE) benchmark. SECURE can comprehensively evaluate LLMs, ensuring they can be trusted as reliable advisors in the high-stakes field of cybersecurity. Below, we illustrate the main components in the design of SECURE.

\subsection{Modeling}\label{sec:modeling}

We emphasize \textit{knowledge modeling} in designing our evaluation benchmark. Knowledge, which includes both facts and skills, is a core indicator of intelligence \cite{feigenbaum1977art}. Prior research shows that knowledge-intensive tasks can reliably evaluate the capabilities of LLMs \cite{yu2023kola, petroni2019language}. For cyber-advisory LLM, we aim to assess not only its ability to retrieve known facts but also gauge its proficiency in a) \textit{using context to answer questions} and b) \textit{performing reasoning based on given knowledge source}. To create a robust evaluation benchmark, we focus on the following critical abilities of LLMs in handling knowledge-intensive tasks:

\textbf{Extraction:} Knowledge extraction tasks are critical for assessing a language model's ability to access and accurately retrieve specific information from its extensive knowledge base \cite{petroni2019language}. In the context of cybersecurity advisory, an LLM may be required to provide information on various security frameworks, incidents, best practices, and historical data on known vulnerabilities. Accurate extraction is essential to ensure the LLM can deliver reliable and precise information promptly, which is crucial for addressing security concerns, preventing breaches, and assisting security professionals in their decision-making processes.

\textbf{Understanding:} While knowledge extraction tasks focus on recalling information, knowledge understanding tasks are designed to assess the cognitive abilities of a model \cite{collins2014knowledge}. This involves evaluating the model's capability to discern the truthfulness of statements and comprehend underlying knowledge within a given context. For instance, an LLM might be tested on its ability to interpret the accuracy of security issues described in a report, thereby demonstrating its grasp of complex cybersecurity concepts and scenarios. Effective understanding ensures that the LLM can accurately interpret and respond to nuanced security challenges.

\textbf{Reasoning:} Knowledge reasoning tasks aim to evaluate the problem-solving capabilities of LLMs \cite{cao2022kqa}. This is particularly important in the cyber-advisory role of LLMs, as they need to assist security professionals in reading, analyzing, and summarizing extensive and detailed threat reports. Effective reasoning enables the LLM to make informed recommendations, identify potential security risks, and suggest mitigation strategies based on the comprehensive analysis of the available data.

\subsection{Data Sources and Tasks}\label{sec:datasourcetask}

Employing a proprietary LLM like ChatGPT-4 \cite{gpt4} is now a research standard in generating evaluation benchmarks in long-form responses \cite{yu2023kola, kim2023prometheus, zheng2024judging}. We utilize the more recent OpenAI's ChatGPT-4o \cite{gpt4o} in extracting our benchmark datasets, using suitable prompts. We also evaluate the quality of ChatGPT-4o output with human annotators to discard incorrect responses (annotation process and results discussed in Section \ref{sec:datasetvalidate}). Below we explain the three different tasks of SECURE and various data sources:

\begin{table*}[]
\caption{Sample MCQs and Answers from MAET and CWET}
\label{tab:mcq-sample}
\centering
\resizebox{\textwidth}{!}{%
\begin{tabular}{@{}lllllll@{}}
\toprule
URL & Question & Option A & Option B & Option C & Option D & Answer \\ \midrule
\href{https://attack.mitre.org/techniques/T0848/}{\textcolor{blue}{Link1}} & \begin{tabular}[c]{@{}l@{}}What is a primary purpose of setting up\\ a rogue master in an ICS environment?\end{tabular} & \begin{tabular}[c]{@{}l@{}}Sending legitimate control \\ messages to devices\end{tabular} & \begin{tabular}[c]{@{}l@{}}Intercepting internal \\ communications\end{tabular} & Encrypting network traffic & \begin{tabular}[c]{@{}l@{}}Creating network \\ segmentation\end{tabular} & A \\
\href{https://attack.mitre.org/mitigations/M0924/}{\textcolor{blue}{Link2}} & \begin{tabular}[c]{@{}l@{}}Which technique is addressed by\\ restricting registry permissions in ICS?\end{tabular} & Phishing & Denial of Service & \begin{tabular}[c]{@{}l@{}}Unauthorized Registry \\ Modifications\end{tabular} & SQL Injection & C \\ \bottomrule
\end{tabular}%
}

\end{table*}

\subsubsection{Extraction Task}\label{sec:extractiontask}
We frame the knowledge extraction task as a multiple-choice question answering (MCQ) task. The model is expected to answer questions without any given context, relying solely on its memory or training data. For this purpose, we focus on the MITRE ATT\&CK~\cite{mitreOrg} and CWE (Common Weakness Enumeration)~\cite{cwe1358} websites to create two datasets: MAET (Mitre Attack Extraction Task) and CWET (Common Weakness Extraction Task). Specifically, we utilize the attack patterns for ICS \cite{mitre-techniques} and mitigation plans for ICS \cite{mitre-mitigations} to generate the questions. For CWE, we utilize the weaknesses belonging to the class CWE-1358 ``CWE VIEW: Weaknesses in SEI ETF Categories of Security Vulnerabilities in ICS'' \cite{cwe1358}, which enumerates all ICS-related security vulnerabilities.

We selected these resources due to their high relevance and quality in the domain of ICS cybersecurity. The MITRE ATT\&CK framework is a globally recognized repository of tactics and techniques based on real-world observations, making it an authoritative source for constructing a benchmark dataset \cite{mitreOrg}. Similarly, the CWE provides a community-developed list of software and hardware weaknesses that can become vulnerabilities, which is crucial for a comprehensive understanding of ICS security. The detailed and structured information from both MITRE ATT\&CK and CWE ensures that the generated questions cover fundamental concepts and advanced technical details pertinent to ICS security. Below is an example prompt used in generating the questions for the MAET and CWET tasks in Prompt A (shortened). Using such prompts, we extract a total of 2036 questions. These questions test both basic and advanced understanding of ICS cybersecurity, ensuring the robustness of SECURE in evaluating the knowledge-intensive tasks by the LLM.

\begin{highlighted}{}
\footnotesize
\underline{Prompt A:} From the following URL `\$\{URL\}', generate a set of MCQs (zero to five) for `novices' and similarly for `experts' with four possible answers each. ...  Return the output in CSV format (tab separated) for the responses with the following nine columns: URL, Level (Novice or Expert), Question, Option A, Option B, Option C, Option D, Correct Answer (A, B, C or D), Explanation...
\end{highlighted}


\underline{\textbf{Use-case:}} Knowledge extraction tasks measure the ability of a language model to access its vast knowledge base and accurately recall specific facts. This is useful when a cybersecurity professional has to find answers to a threat, system weakness, mitigation plans, attack patterns, and vulnerability given a specific attack scenario or just simply to gain information to assist the other tasks they are performing. LLMs are considered to be repositories of world knowledge with few-shot learning \cite{brown2020language} and zero-shot reasoning \cite{kojima2022large} skills and hence are frequently inquired to obtain answers for different questions. MAET and CWET, both extraction tasks, consist of such questions that replicate this situation and evaluate if LLMs have sufficient knowledge in the ICS domain. Since state-of-the-art LLMs are trained on world knowledge and have seen documents on MITRE and CWE, such questions measure whether they can accurately recall answers to these kinds of questions.

\subsubsection{Understanding Task}\label{sec:understandingtask}
Our second task is designed to evaluate the ability of LLM to comprehend and understand the security-related text. Given the continuously evolving landscape of cybersecurity, it is crucial that LLMs can assimilate new information and use it to generate accurate responses. For this purpose, we utilize the CVEs (Common Vulnerabilities and Exposures) published in 2024, available at CVE project repository \cite{cve2024}. CVE is better for this task since they are always updated but CWEs are more appropriate for the extraction task as they are updated less frequently.

To ensure the integrity of our evaluation, we verified that none of the pretrained models we selected for evaluation (ChatGPT-4 \cite{gpt4}, ChatGPT-3.5 \cite{gpt3.5}, Llama3-70b \cite{llama370b}, Llama3-8b \cite{llama38b}, Gemini-Pro \cite{gemini}, Mistral-7B \cite{jiang2023mistral} , Mixtral-8x7b \cite{mixtral8x7b}) had access to these CVEs during their training phase by confirming their training-cut-off date (See Table \ref{tab:LLMCompare}). This precaution guarantees that the models have not been exposed to this specific information and thus must rely on their comprehension abilities, and therefore, maintain the validity and reliability of SECURE.

We generate the KCV (Knowledge test on Common Vulnerabilities ) dataset, a series of boolean questions that require the LLMs to read the CVE descriptions provided in JSON format and determine whether the given statements are True or False based on the available information. This setup tests the models' ability to accurately process and understand newly introduced data.

In addition, we created a supplementary dataset named VOOD (Vulnerability Out-of-Distribution task), which contains questions without relevant context to assess the model's ability to recognize when it lacks sufficient information to answer a question. None of these questions could be answered truthfully without access to the discussed vulnerability. Ideally, the model should indicate its inability to answer in such cases.

For both KCV and VOOD, we generated 466 boolean questions. These datasets serve as a robust benchmark for evaluating the comprehension capabilities of LLMs in cybersecurity, mainly focusing on their ability to adapt to and reason about newly encountered vulnerabilities.

\underline{\textbf{Use-case:}} Knowledge understanding tasks are designed to assess the model’s capability to discern the truthfulness of statements with and without a context. KCV evaluates whether existing LLMs can comprehend security documents, and answer questions based on a given context. VOOD inspects how the models perform when the context is not provided. Both of these tasks measure the reliability of LLMs. An effective understanding of cybersecurity can only ensure that the LLM can accurately interpret and respond to nuanced security challenges. 

\begin{table*}[]
\caption{An example row from RERT dataset. Overview \& vulnerability description shortened for readability.}
\label{tab:reasondataset}
\resizebox{\textwidth}{!}{%
\begin{tabular}{@{}llll@{}}
\toprule
\textbf{URL} & \textbf{Overview} & \textbf{Vulnerability} & \textbf{Risk Evaluation} \\ \midrule
\href{https://www.cisa.gov/news-events/ics-advisories/icsa-24-142-01}{\textcolor{blue}{Link}} & \begin{tabular}[c]{@{}l@{}}Successful exploitation of this vulnerability could allow an attacker \\ to inject arbitrary JavaScript into a user's web browser for a single \\ vulnerability, successful exploitation of these vulnerabilities could \\ cause a denial of service, disclosure of sensitive information, \\ communication loss, and modification of settings or ladder logic \\ for multiple vulnerabilities.\end{tabular} & \begin{tabular}[c]{@{}l@{}}3.2.1 Path Traversal CWE-22 There are multiple ways in LAquis \\ SCADA for an attacker to access locations outside of their own \\ directory. CVE-2024-5040 has been assigned to this vulnerability. \\ A CVSS v3.1 base score of 7.8 has been calculated;\end{tabular} & \begin{tabular}[c]{@{}l@{}}Successful exploitation of this vulnerability could \\ allow an attacker to read and write files outside \\ of their own directory.\end{tabular} \\ \bottomrule
\end{tabular}%
}

\end{table*}

\subsubsection{Reasoning Task}\label{sec:reasoningtask}
Our third task evaluates the LLM's reasoning capability in the context of ICS security through the Risk Evaluation Reasoning Task (RERT). To create RERT, we compile a risk assessment dataset using cybersecurity advisories from the Cybersecurity and Infrastructure Security Agency (CISA) \cite{cisa_advisories}. We focus on all significant ICS-related advisory reports, each containing an \textit{Executive Summary}, \textit{Risk Evaluation}, \textit{Technical Details} (including \textit{Vulnerability Overview}, \textit{Affected Products}, \textit{Background Researcher}), and \textit{Mitigations}. Reports lacking essential information are discarded to ensure data quality, and the remaining comprehensive reports are processed for inclusion.

The \textit{Risk Evaluation} section of these advisories consistently summarizes the key risks associated with the identified vulnerabilities. This summary is presented in a specific format that can be inferred from the detailed vulnerability information, assuming a sufficient background in cybersecurity. As these reports are meticulously curated by security professionals, we treat their risk evaluations as gold standards, allowing us to accurately assess the performance of LLMs against the high benchmarks set by human experts.

Therefore, the task involves predicting the \textit{Risk Evaluation} based on the provided vulnerability details. This setup allows us to gauge the LLM's ability to understand and reason about complex ICS security scenarios. We compiled 1,000 samples from the most recent ICS advisories to construct this dataset, ensuring a robust and up-to-date benchmark for evaluation.

In addition, we create another dataset CPST (CVSS Problem Solving Task) to measure the problem-solving skills of LLMs in security. Specifically, we use the CVSS (Common Vulnerability Scoring System) framework and manually collect 100 unique CVSS3.1 vector strings from the Cybersecurity and Infrastructure Security Agency (CISA) \cite{cisa_advisories}. These scores, in the range of 0-10, can be computed using the CVSS calculator which uses the Base, Temporal, and Environmental scores to determine the overall severity of a vulnerability \cite{cvss3}. This task involves using an existing formula from CVSS3.1 standard and computing a value. This computation allows the evaluation of the problem-solving skills of LLMs in practical security settings.

\underline{\textbf{Use-case:}} LLMs are now integrated into emails and document software to summarize conversations and texts. Such LLMs work quite well in handling the general English language but summarizing a threat report is significantly different as the LLMs need to understand the technical details that consist of vulnerability, affected products, and risks. RERT dataset is designed to evaluate the ability of LLMs in summarizing an extensive threat report. The CVSS Problem Solving Task checks if LLMs can understand and use the CVSS formula without being explicitly programmed (zero-shot evaluation). While a simple program can do this deterministically, this task shows if LLMs can help in real-world scenarios where users might not know the formula. It shows LLMs' ability to reason and solve problems, not just follow set rules.

\subsection{Dataset Validation}\label{sec:datasetvalidate}
To ensure the quality and validity of our evaluation datasets, we conducted a rigorous manual verification process for all generated questions. This process involved independent assessment by human annotators with expertise in computer security, specifically Masters or Ph.D. students in the field. Discrepancies between the human annotators' labels and the original ground truth provided by ChatGPT-4o were resolved through adjudication by a second, more experienced annotator.

Our analysis revealed several categories of issues within the generated questions. Some questions were deemed unanswerable from the provided context, particularly within the True/False format. In the Multiple Choice Question (MCQ) format, we encountered instances where multiple answer choices were deemed correct. To maintain the integrity of our evaluation, we removed questions deemed unanswerable or exhibiting multiple correct answers from the dataset. Questions with identifiable issues that could be rectified were corrected accordingly. We fixed 13 questions in MAET, 0 in CWE and 22 questions in KCV. This manual verification and refinement process removed a small percentage of questions originally collected for each dataset. Specifically, we removed 2.9\% of questions from the MAET dataset, 3.5\% from the CWET dataset, and 6.9\% from the KCV dataset, totaling 29, 36, and 32 questions respectively.

\subsection{Benchmark Dataset and Evaluation}
Based on the modeling and task descriptions of Section \ref{sec:modeling} and \ref{sec:datasourcetask}, we have created the following benchmark datasets:

\textbf{Knowledge extraction dataset:} As detailed in Section \ref{sec:extractiontask}, these tasks are framed as MCQs derived from MITRE ATT\&CK and CWE websites. We have released two datasets: \textbf{MAET} and \textbf{CWET}, comprising a total of 2036 MCQs. We show a sample in Table \ref{tab:mcq-sample}. Different LLMs are evaluated based on their accuracy in predicting the correct answers from the provided options, allowing us to measure their effectiveness in knowledge extraction.

\begin{table}[]
\caption{Sample from KCV and VOOD dataset}
\label{tab:bool-sample}
\centering
\resizebox{0.45\textwidth}{!}{%
\begin{tabular}{@{}llr@{}}
\toprule
\textbf{CVE-ID} & \textbf{Question} & \multicolumn{1}{l}{\textbf{Answer}} \\ \midrule
CVE-2024-0011 & \begin{tabular}[c]{@{}l@{}}The vulnerability described in CVE-2024-0011 allows \\ for the execution of arbitrary code on the affected system.\end{tabular} & FALSE \\ \midrule
CVE-2024-0017 & \begin{tabular}[c]{@{}l@{}}User interaction is required for the exploitation of the \\ CVE-2024-0017 vulnerability.\end{tabular} & TRUE \\ \bottomrule
\end{tabular}%
}
\end{table}

\textbf{Knowledge understanding dataset:} As explained in Section \ref{sec:understandingtask}, we utilize CVE published in 2024 to create this boolean dataset. There are two variants:~\textbf{KCV}, which includes the context from CVE JSON files, and \textbf{VOOD}, which lacks this context to assess the out-of-distribution performance of LLMs. Each dataset contains 466 Boolean questions. Along with the questions and answers, we also provide the JSON files used as context for evaluating the LLM's comprehension abilities. A sample is shown in Table \ref{tab:bool-sample}. Different LLM models are evaluated based on their accuracy in predicting the truthfulness of statements, both with and without context.

\begin{table}[]
\caption{Sample from CPST dataset}
\label{tab:cvss-table}

\centering
\resizebox{0.35\textwidth}{!}{%
\tiny 
\begin{tabular}{@{}l c @{}}
\toprule
\textbf{CVSS v3 Vector String} & \textbf{Correct Answer} \\ \midrule
AV:L/AC:L/PR:N/UI:R/S:U/C:H/I:H/A:H & 7.8 \\ 
AV:N/AC:H/PR:N/UI:R/S:C/C:L/I:L/A:N & 4.7 \\
AV:N/AC:L/PR:N/UI:N/S:U/C:H/I:H/A:H & 9.8 \\ \bottomrule
\end{tabular}%
}

\end{table}

\textbf{Knowledge reasoning dataset:} We have released \textbf{RERT}, a dataset consisting of 1000 questions based on security advisories from CISA, as detailed in Section~\ref{sec:reasoningtask}. We show a sample in Table \ref{tab:reasondataset}. Different LLM models are evaluated using the ROGUE-L metric \cite{lin-2004-rouge} between the LLM-generated response and the ground truth. Additionally, we have released \textbf{CPST}, which includes 100 manually crafted CVSS3.1 vector strings along with their associated vulnerability scores. A sample of these is provided in Table~\ref{tab:cvss-table}. LLM models are evaluated using the mean average deviation (MAD) against the ground-truth scores, assessing their accuracy in generating precise vulnerability assessments.

\section{Experiments \& Results}\label{sec:experiments}

We evaluate 7 state-of-the-art LLMs varying in parameter size, organization, and access (See Appendix \ref{appendix:models} and Table \ref{tab:LLMCompare} for more detail). We pick both open-source and closed models based on LLM leaderboards (ChatArena \cite{ChatArena}). Open-source models allow the download of full model weights whereas closed models provide an API for restricted access. We evaluate our benchmark against the following models:

\textbf{Open-source models:} Llama3-70B \cite{llama370b}, Llama3-8B \cite{llama38b}, Mistral-7B \cite{jiang2023mistral}, Mixtral-8x7b \cite{mixtral8x7b}

\textbf{Closed-source models:} ChatGPT-3.5 \cite{gpt3.5}, ChatGPT-4 \cite{gpt4}, Gemini-Pro (v1.5) \cite{gemini}

\begin{table*}[]
\centering
\caption{Result of different models on our benchmark tasks. $\uparrow$-$\downarrow$ indicate higher-lower values are better.}
\resizebox{0.85\textwidth}{!}{%
\begin{tabular}{@{}lcccccc@{}}
\toprule
\textbf{Model} & \textbf{MAET (Acc $\uparrow$)} & \textbf{CWET (Acc $\uparrow$)} & \textbf{KCV (Acc$\uparrow$)} & \textbf{VOOD (OOD-Acc$\uparrow$)} & \textbf{RERT (ROGUE-L$\uparrow$)} & \textbf{CPST (MAD$\downarrow$)} \\ \midrule
ChatGPT-4 & \textbf{88.6} & \textit{89.6} & \textbf{87.6} & \textbf{87.9} & 0.53 & \textbf{0.81} \\
ChatGPT-3.5 & 82.8 & 84.2 & 78.3 & 8.4 & 0.48 & 1.26 \\
Gemini-Pro & \textit{86.2} & 87.8 & 83.5 & 86.7 & \textbf{0.54} & \textit{1.0} \\ \hline
Llama3-70B & 86.3 & \textbf{90.4} & \textit{85.2} & 27.1 & 0.51 & 1.54 \\
Llama3-8B & 82.1 & 83.9 & 82.8 & 56.4 & 0.48 & 1.77 \\
Mistral-7B & 77.9 & 80.1 & 64.2 & 57.1 & 0.42 & 1.82 \\
Mixtral-8x7B & 80.9 & 83.4 & 79.6 & 69.3 & 0.39 & 1.63 \\ \bottomrule
\end{tabular}%
}

\label{tab:results}
\end{table*}

\subsection{Prompting Strategy for Evaluation}
We use consistent prompt engineering approach to ensure uniformity across different LLMs. Instead of customizing prompts for each model, we employed a single prompt structure per task type that produced responses in the required format. 

\textbf{Example Prompt (MCQ Tasks: MAET and CWET)}: \code{ For the given question: Which protocol function can be disabled to prevent unauthorized device shutdowns?, and four options: A) DNP3 0x0D, B) HTTP GET, C) SMTP HELO, or D) FTP LIST, pick the best option as the answer, and return as either A, B, C or D. If you do not know the answer, return X. Choose the appropriate letter from A, B, C, D, or X as your answer. Please provide only the letter corresponding to your choice without any additional text or explanations.}

We embedded relevant information from the JSON files directly within the prompt for the KCV task, which requires external knowledge. To maintain compatibility with model input constraints and avoid exceeding maximum token limits, we selectively included questions whose prompt length, including the embedded knowledge, did not exceed 5000 words. Although the models generally followed the specified output format, there were instances where responses deviated slightly from the instructions. In such cases, we manually corrected the outputs for accurate evaluation. We include all prompts we used to generate the responses in our dataset for reproducibility, and demonstrate more examples in Appendix \ref{appendix:prompt}.

\textbf{Temperature Setting}: LLM temperature influences the output of language models by adjusting the randomness or predictability of generated text. A higher temperature results in more creative but potentially less coherent outputs, as the model is more likely to choose less probable words. Conversely, a lower temperature makes the output more deterministic and predictable, often resulting in repetitive and conservative responses. Typically set between 0 and 1, the temperature modifies the probability distribution of the next word in a sequence. We use the default temperature parameter (set at 0.7) for all evaluation prompts.

\subsection{Evaluation Metrics}
We employ a range of evaluation metrics tailored to the specific nature of each task within our benchmarking framework: accuracy for \textbf{MAET}, \textbf{CWET}, \textbf{VOOD} and \textbf{KCV}, ROGUE-L \cite{lin-2004-rouge} for \textbf{RERT} and mean absolute deviation (MAD) for \textbf{CPST}. We explain the metrics in detail in Appendix \ref{appendix:evalmetrics}.

\subsection{Results Summary}

Table \ref{tab:results} presents a comparative summary of the performance of various language models (LLMs) evaluated across the six benchmark tasks of SECURE. Our analysis reveals a consistent trend of closed-source models, particularly ChatGPT-4 and Gemini-Pro, exhibiting superior performance across most tasks.

ChatGPT-4 emerges as the top performer, achieving the highest scores on four key metrics: accuracy in MAET (88.6\%), KCV (87.6\%), and VOOD (87.9\%), as well as the lowest mean absolute deviation on CPST (0.81). These results underscore ChatGPT-4's robust capabilities in handling a variety of tasks, including out-of-distribution detection and complex problem-solving scenarios. Gemini-Pro closely trails ChatGPT-4, achieving comparable results on most tasks and slightly outperforming it on the summarization task based on the ROUGE-L metric. While ChatGPT-3.5 shows decent overall performance, it struggles significantly on the VOOD task, with an accuracy of only 8.4\%.

Among the open-source models evaluated, Llama3-70B stands out due to its higher model capacity, consistently demonstrating superior performance. It achieves results comparable to ChatGPT-4 and even surpasses it on the CWET task. However, its subpar performance on the VOOD task indicates a potential weakness in handling out-of-domain scenarios. In contrast, models such as Llama3-8B, Mistral-7B, and Mixtral-8x7B exhibit moderate performance levels across the evaluated tasks.

\section{Discussion \& Analysis}\label{sec:resultsAnalysis}

\subsection{Error Analysis} 
\label{sec:err-analysis}

In this section, we analyze the incorrect responses of LLMs on some representative examples to understand their shortcomings.

Q1. \code{What is the primary purpose of setting up a rogue master in an ICS environment? A) Sending legitimate control messages to devices, B) Intercepting internal communications,	C) Encrypting network traffic, and D) Creating network segmentation.}

The correct answer for this question is \textit{A) Sending legitimate control messages to devices}. However, among all the LLMs evaluated, only ChatGPT-4 provided the correct response. This discrepancy might be due to the LLMs difficulty in recognizing the malicious intent embedded in the term ``legitimate control messages''. The positive sentiment associated with the word ``legitimate'' could have misled the models, causing them to overlook the actual malicious purpose.

Q2. \code{What is the main vulnerability exploited in CAPEC-477?	A) Incorrect hashing algorithm, B) Incorrect data storage structure, C) Mixing signed and unsigned content, D) Weak encryption}

The correct answer for this question is \textit{C) Mixing signed and unsigned content}. Only ChatGPT-4 and Llama3-70B managed to answer correctly. This might be because options A, B, and D explicitly mention common vulnerability types, which could have led the models to favor these options over the correct one. The subtlety of mixing signed and unsigned content as a specific vulnerability might have been less recognizable to the other LLMs.

Q3. \code{State whether this statement is True or False given the JSON file as context (Source\footnote{\url{https://github.com/CVEProject/cvelistV5/blob/main/cves/2024/36xxx/CVE-2024-36039.json}}): The CVE-2024-36039 vulnerability is caused by improper escaping of JSON values when using PyMySQL.	Answer: F}

All LLMs incorrectly answered this statement as True. The JSON file clearly states that the vulnerability is due to improper escaping of JSON keys, not values. This indicates that the models failed to differentiate between JSON values and keys within the given context. Such nuanced distinctions are crucial for accurate comprehension and response, highlighting a significant area for improvement in the models' contextual understanding and attention to detail.

\subsection{Impact of Confidence on LLM Accuracy}

Prior studies have demonstrated that large language models (LLMs) can be calibrated through self-reflection or confidence analysis \cite{chen2023quantifying}. In this study, we evaluate the relationship between model confidence and performance across different confidence levels using the CWET task. Specifically, we measure the confidence of two representative LLMs, ChatGPT-4 and Llama3-70B, across five temperature settings: 0.6, 0.7, 0.8, 0.9, and 1.0, to obtain a more robust confidence estimate.

\begin{figure}[]
\centering
  \centering
\includegraphics[width=0.84\linewidth]{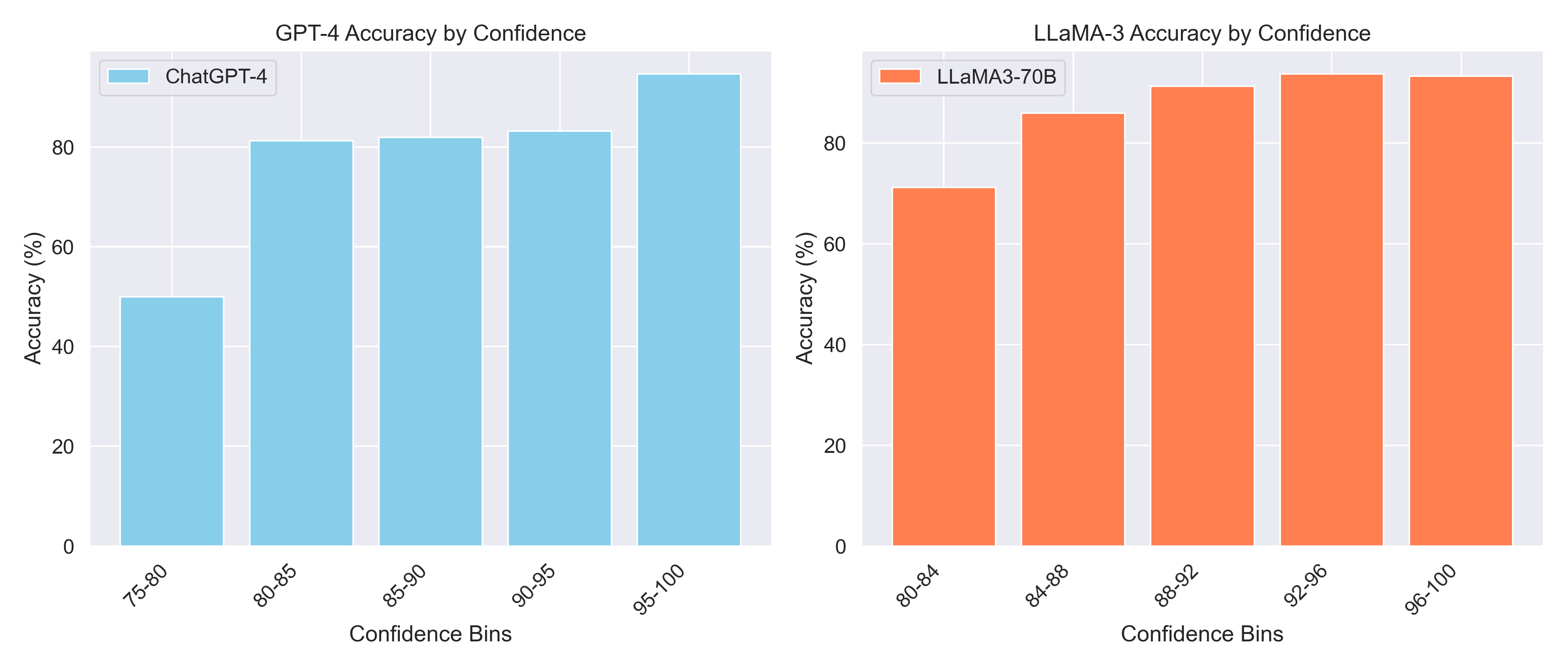}
    \caption{Confidence vs. Accuracy of LLMs}
    \label{fig:confidence}

\end{figure}

The LLMs were prompted to provide both their answers and the probability that their answers were correct (ranging from 0\% to 100\%), formatted as follows: \code{``Provide your answer and the probability that the answer is correct (0\% to 100\%) separated by a space.''} We then averaged the confidence scores provided by the LLMs across the different temperatures. The accuracy of the models was plotted against five different confidence bins, as shown in Figure~\ref{fig:confidence}.

The results indicate a clear trend: as confidence decreases, so does accuracy, particularly within the lowest confidence bins. Furthermore, ChatGPT-4 consistently exhibits higher confidence scores than Llama3.
These findings suggest that different LLMs may necessitate tailored calibration techniques to effectively mitigate incorrect responses. Specifically, while both models show a correlation between confidence and accuracy, the disparity in their confidence levels highlights the need for model-specific calibration. Future research could explore more sophisticated self-reflection and confidence estimation techniques to further improve the reliability and accuracy of LLMs in the cyber advisory tasks.

\begin{highlighted}{}
\small
\textbf{Finding: } Responses from various LLMs with lower confidence levels tend to be less accurate, and different LLMs exhibit varying degrees of confidence, impacting their overall reliability.
\end{highlighted}

\subsection{Open vs. Closed Model Performance}

\begin{figure}[]
  \centering    \includegraphics[width=0.60\linewidth]{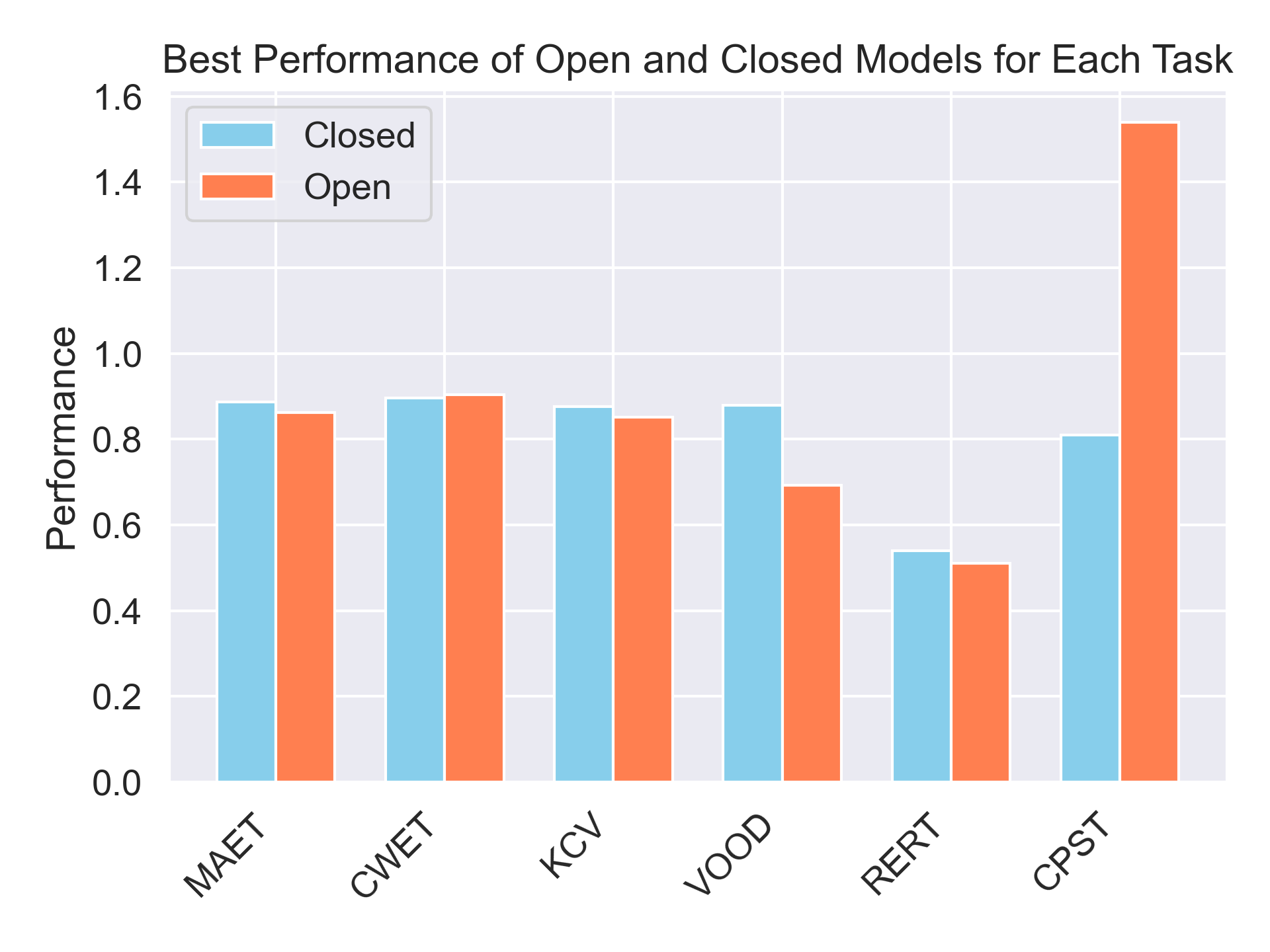} 
    \caption{Performance Comparison of Open-Source and Closed-Source LLMs Across Different Tasks (Note: For CPST, lower scores indicate better performance).}
    \label{fig:open-closed}
\end{figure}

This section analyzes the performance disparity between open-source and closed-source LLMs utilized in our study. Figure~\ref{fig:open-closed} presents a comparative analysis showcasing the peak performance achieved by the three closed LLMs (ChatGPT-3.5, ChatGPT-4, and Gemini-Pro) and the remaining four open-source LLMs. 

Our analysis reveals a consistent trend of closed-source models outperforming their open-source counterparts across most tasks. ChatGPT-4 demonstrates superior performance across all tasks except CWET, where Llama3-70B achieves marginally better results. While the performance difference remains relatively small for MAET, CWET, KCV, and RERT tasks, it becomes more pronounced in VOOD and CPST.

The superior performance of ChatGPT-4 and Gemini-Pro on VOOD (Out-of-Distribution Detection) suggests the implementation of more robust safeguarding mechanisms within these models. This observation is particularly evident in the substantial performance difference between ChatGPT-4 and ChatGPT-3.5, highlighting significant advancements in out-of-distribution detection capabilities. Closed models also exhibit significantly better performance on the CPST task, indicating a higher proficiency in problem-solving related to CVSS score calculation.

However, these safeguarding features, while crucial for enhancing security, may inadvertently hinder analysis by restricting access to crucial cyber threat intelligence. This limitation was particularly pronounced when working with the Gemini model. Specifically, during the RERT summarization task, the API refused 319 out of 1000 queries, deeming them potentially harmful due to their inclusion of vulnerability information. This finding raises a critical question: how can we strike a balance between robust security measures and the necessary access to sensitive information for effective cybersecurity research and practice?

\begin{highlighted}{}
\textbf{Finding: } The performance difference between open and closed models is negligible except for problem-solving or out-of-distribution tasks, where closed LLMs yield better results due to more strict safeguarding.
\end{highlighted}

\subsection{Evaluating Reasoning Abilities}

\begin{figure}[]
  \centering    
  \includegraphics[width=0.75\linewidth]{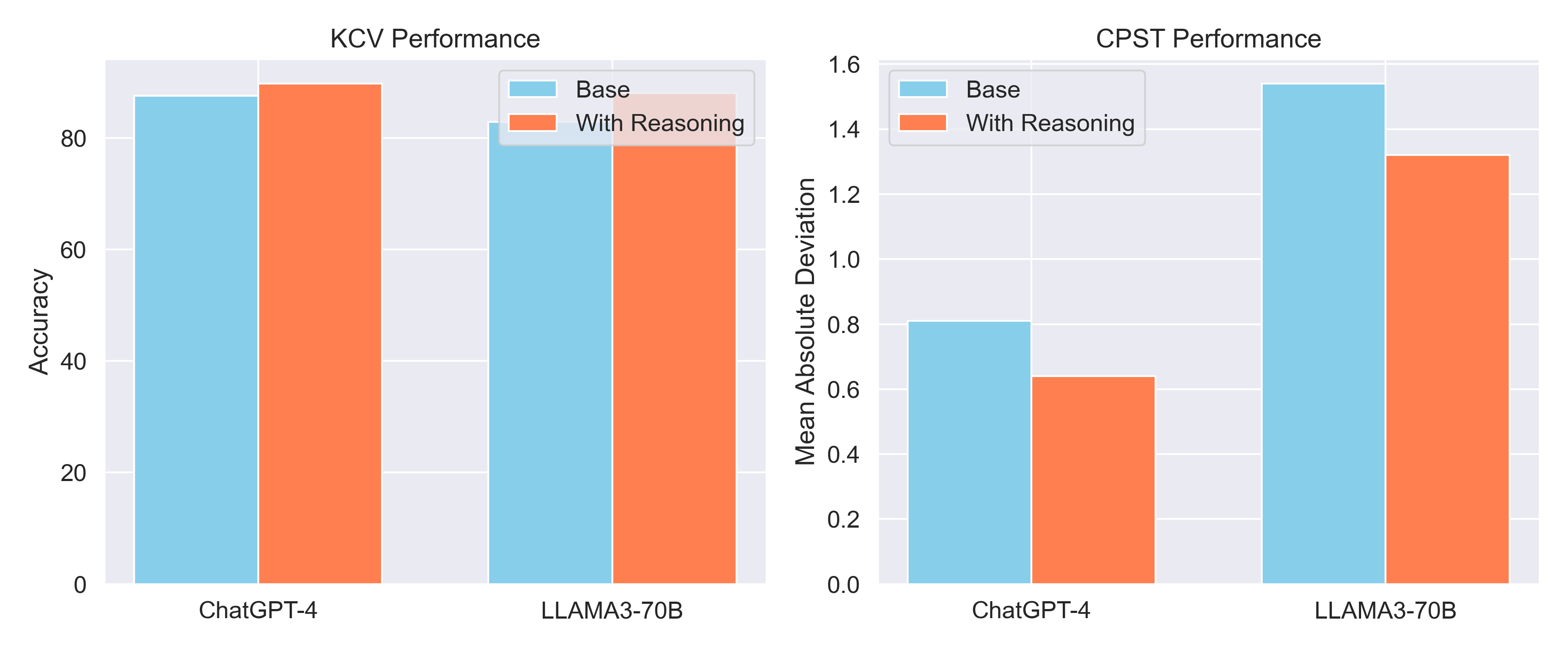} 
    \caption{Performance of LLMs when asked to perform step-by-step analysis}
    \label{fig:reasoning}
\end{figure}

\begin{figure}[]
  \centering    \includegraphics[width=0.75\linewidth]{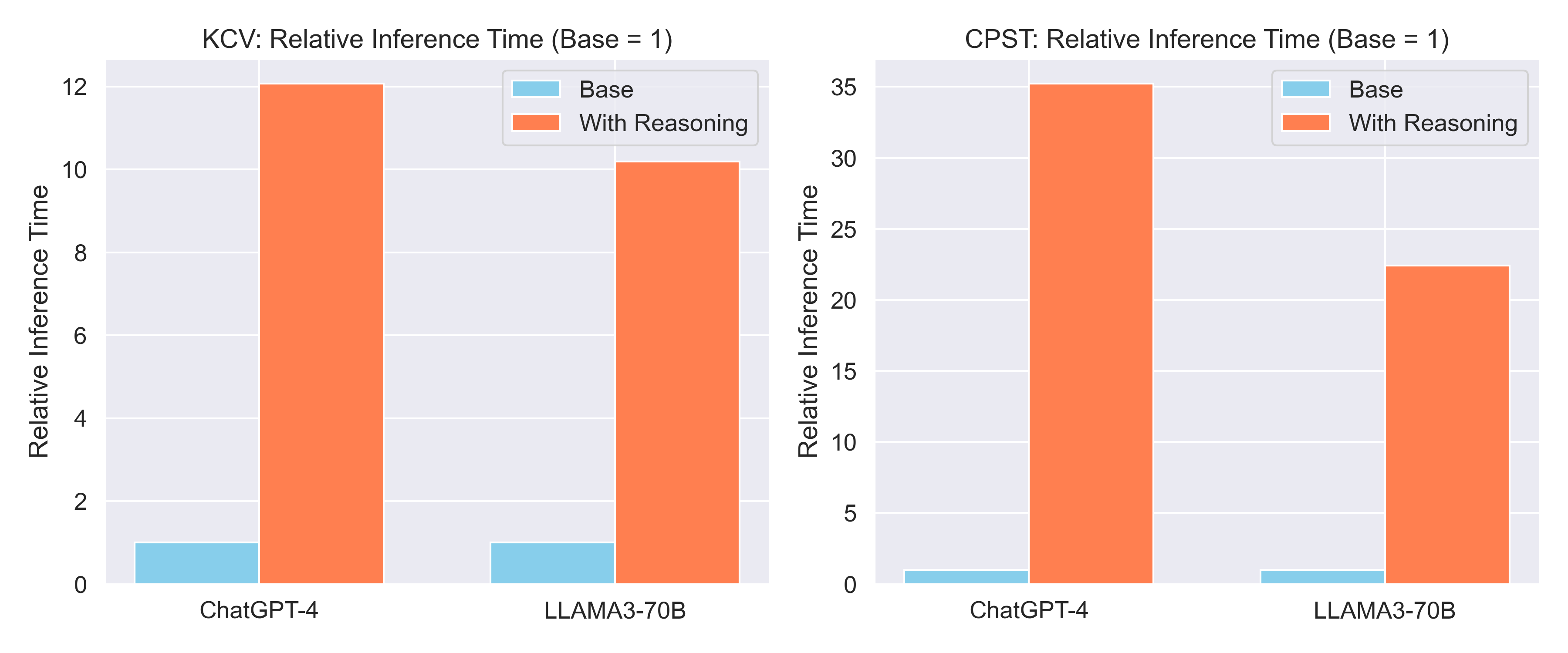} 
    \caption{Inference time on reasoning tasks when asked to perform step-by-step analysis}
    \label{fig:timing}
\end{figure}

We further evaluate LLMs' reasoning capabilities on KCV and CPST by asking the model to perform step-by-step analysis to arrive at the result. For both tasks, we modify the prompt to include the sentence: \code{Provide a detailed explanation of how you arrived at the answer.} 

We assess the performance of the ChatGPT-4 and Llama3-70B models on these tasks. The performance results are illustrated in Figure~\ref{fig:reasoning}. As shown, explicit instructions for providing explanations enhance performance on both tasks for these models. For the KCV task, the relative improvement in accuracy with reasoning is 2.52\% for ChatGPT-4 and 4.43\% for Llama3-70B.  For the CPST task, the relative improvement in the Mean Absolute Deviation (MAD) score with reasoning is 20.99\% for ChatGPT-4 and 14.29\% for Llama3-70B compared to the baseline.

However, this advantage comes with a caveat. Requiring models to provide detailed reasoning steps can increase inference time, as illustrated in Figure~\ref{fig:timing}. The figure demonstrates that the time increase can be substantial depending on the specific task, as much as 35 times compared to the base task for CPST. This potential for elevated inference time translates to additional computational costs, highlighting a crucial trade-off between enhanced reasoning capabilities and computational expenses.

\begin{highlighted}{}
\textbf{Finding: } Explicitly asking LLMs for explanations or details about their reasoning steps to solve a task can yield significant performance gains but usually comes at a higher computational budget.
\end{highlighted}

\subsection{Variance in Prediction}
\begin{figure}[]
  \centering
\includegraphics[width=0.65\linewidth]{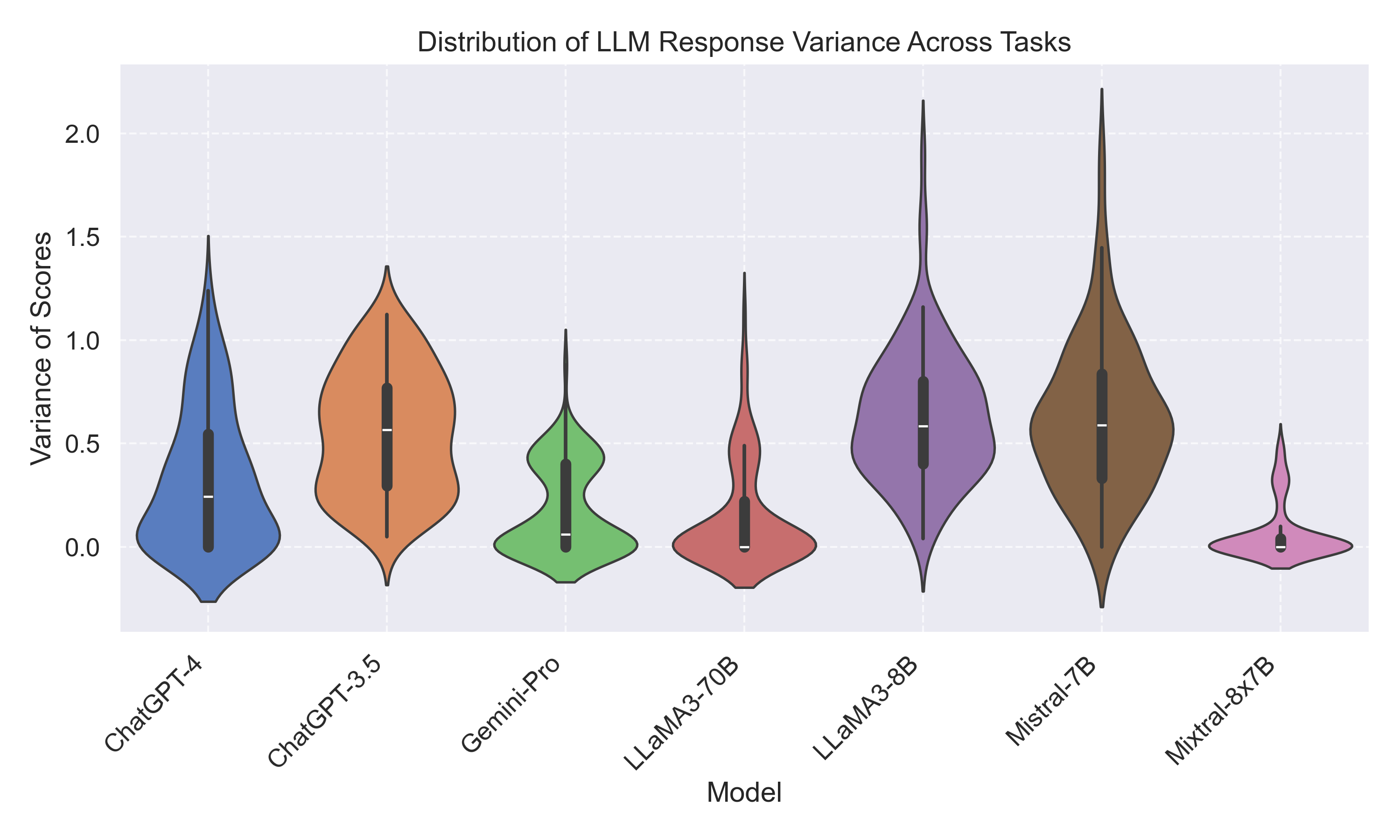}
    \caption{Distribution of variance among predictions for different LLMs}
  \label{fig:variance} 
\end{figure}

Figure \ref{fig:variance} presents a comparative analysis of response variance across different LLMs on the CPST task. This task, requiring a real-number output, is well-suited for distribution analysis. We evaluated each model's responses over five runs, utilizing their default temperature settings. The standard deviation of scores across these runs was calculated for each model, and the resulting distributions are visualized in the violin plot. The visualization reveals a trend: smaller models, such as Llama3-8b and Mistral-7b, tend towards wider variance, suggesting more significant variability in their responses. An exception is Mixtral-8x7b, which exhibits more deterministic behavior despite its smaller size. Larger models generally exhibit minor variance, indicating more stable outputs. However, ChatGPT-4 displays a slightly higher variance than Gemini-Pro and Llama3-70b on this specific task.

\subsection{Model Agreement Bias}

\begin{figure}[]
  \centering
  \begin{subfigure}[b]{0.23\textwidth}
    \includegraphics[width=\textwidth]{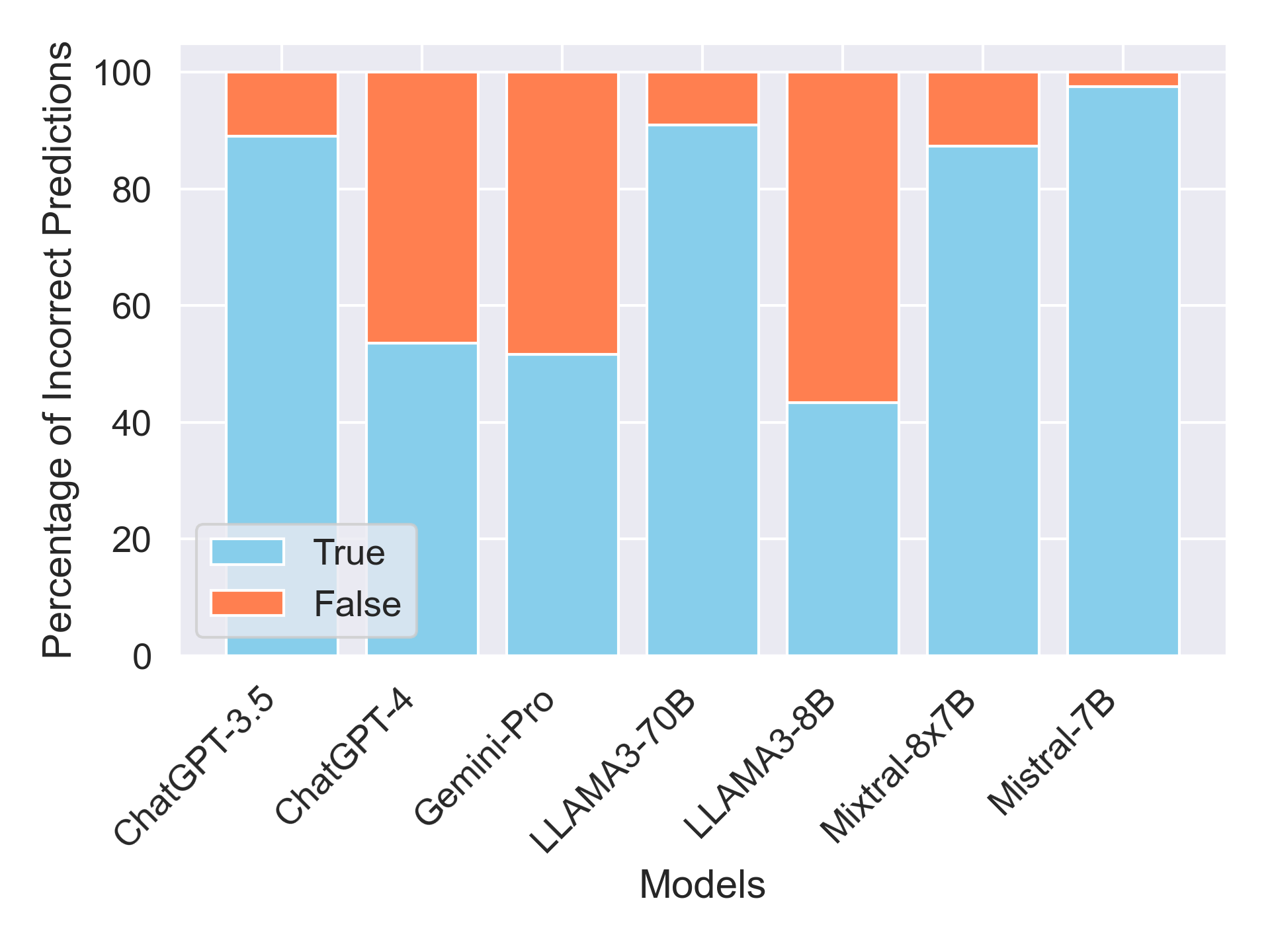}
    \caption{No context}
  \end{subfigure}
  \begin{subfigure}[b]{0.23\textwidth}
    \includegraphics[width=\textwidth]{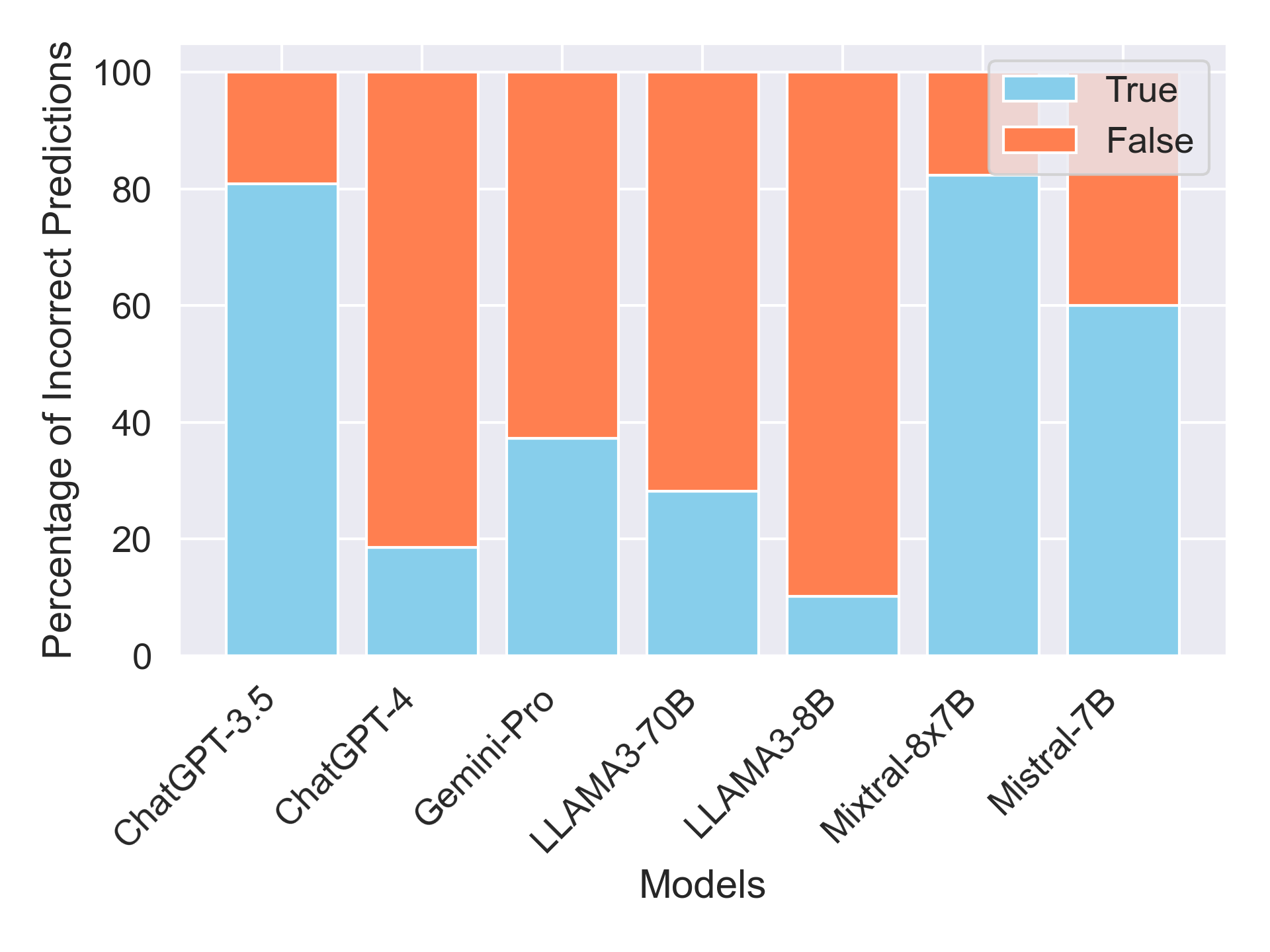}
    \caption{With context}
  \end{subfigure}
  \caption{Fraction of incorrect prediction with no context and context}
  \label{fig:bias}
\end{figure}

This section investigates the tendency of LLMs to exhibit agreement bias when presented with factual statements within the context of cyber advisory. Specifically, we analyze the KCV and VOOD tasks, both of which assess the model's ability to correctly classify a given statement as true or false. The distinction between KCV and VOOD lies in the provision of contextual information. While both tasks utilize the same set of statements, KCV provides additional context relevant to the statement, whereas VOOD presents the statements in isolation. This design allows us to examine how the presence or absence of context influences the model's susceptibility to agreement bias.

Figure~\ref{fig:bias} illustrates the distribution of incorrect predictions for both tasks, categorized by whether the LLM agreed (predicted True) or disagreed (predicted False) with the statement. In VOOD, most errors stem from the LLMs agreeing with the statement, even though the statements related to vulnerabilities were not in their training data. This tendency to affirm novel information suggests a potential for hallucination, where the model generates plausible-sounding but unsubstantiated claims \cite{adlakha2024evaluating}. It is also possible that the models are incorrectly associating information from known CVEs to newer, unseen vulnerabilities.

Conversely, the KCV task, where context is provided, exhibits a higher proportion of errors arising from the LLMs disagreeing with the statement. This pattern indicates a potential limitation in the models' ability to effectively leverage the provided context for accurate information retrieval and association.

\begin{highlighted}{}
\small
\textbf{Finding: } LLMs can hallucinate information when they lack up-to-date information and may fail to infer correct responses even when context is included.
\end{highlighted}

\subsection{Task Correlation Analysis}

\begin{figure}[]
  \centering
\includegraphics[width=0.65\linewidth]{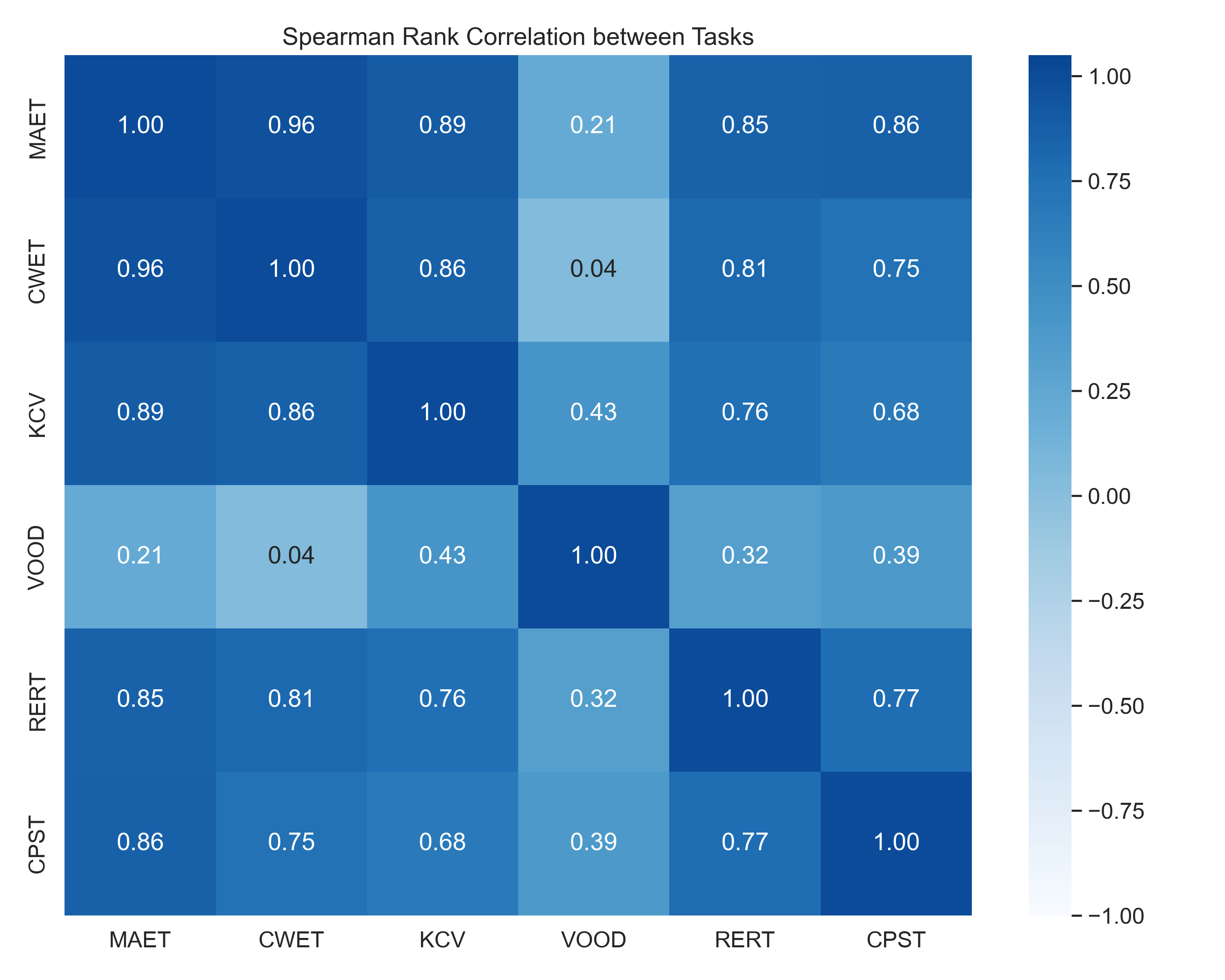}
    \caption{Spearman correlation between different tasks in SECURE benchmark}
  \label{fig:correlation} 
\end{figure}

To understand the relationships between the different tasks in the SECURE benchmark, we conducted a Spearman rank correlation analysis. The results, visualized in Figure \ref{fig:correlation}, reveal interesting insights into the inter-task dependencies. The heatmap shows that most tasks exhibit a moderate to strong positive correlation revealing some inherent relationship between knowledge required for different tasks. For example, the knowledge extraction tasks (MAET) shows notable correlations with other reasoning tasks (CPST, RERT) and comprehension tasks (KCV). This indicates that high-level tasks like reasoning and comprehension also rely on memorized knowledge acquired during model training. As expected, the lowest correlation is between VOOD and other tasks as VOOD task is crafted to evaluate out-of-distribution capabilities of LLMs.

\begin{highlighted}{}
\small
\textbf{Finding: } Different tasks in SECURE exhibit strong inter-task correlations, suggesting that performance gains in one task may also beneficially impact others.
\end{highlighted}

\subsection{LLM Performance Across Expertise Levels}

\begin{figure}[]
  \centering
    \includegraphics[width=0.7\linewidth]{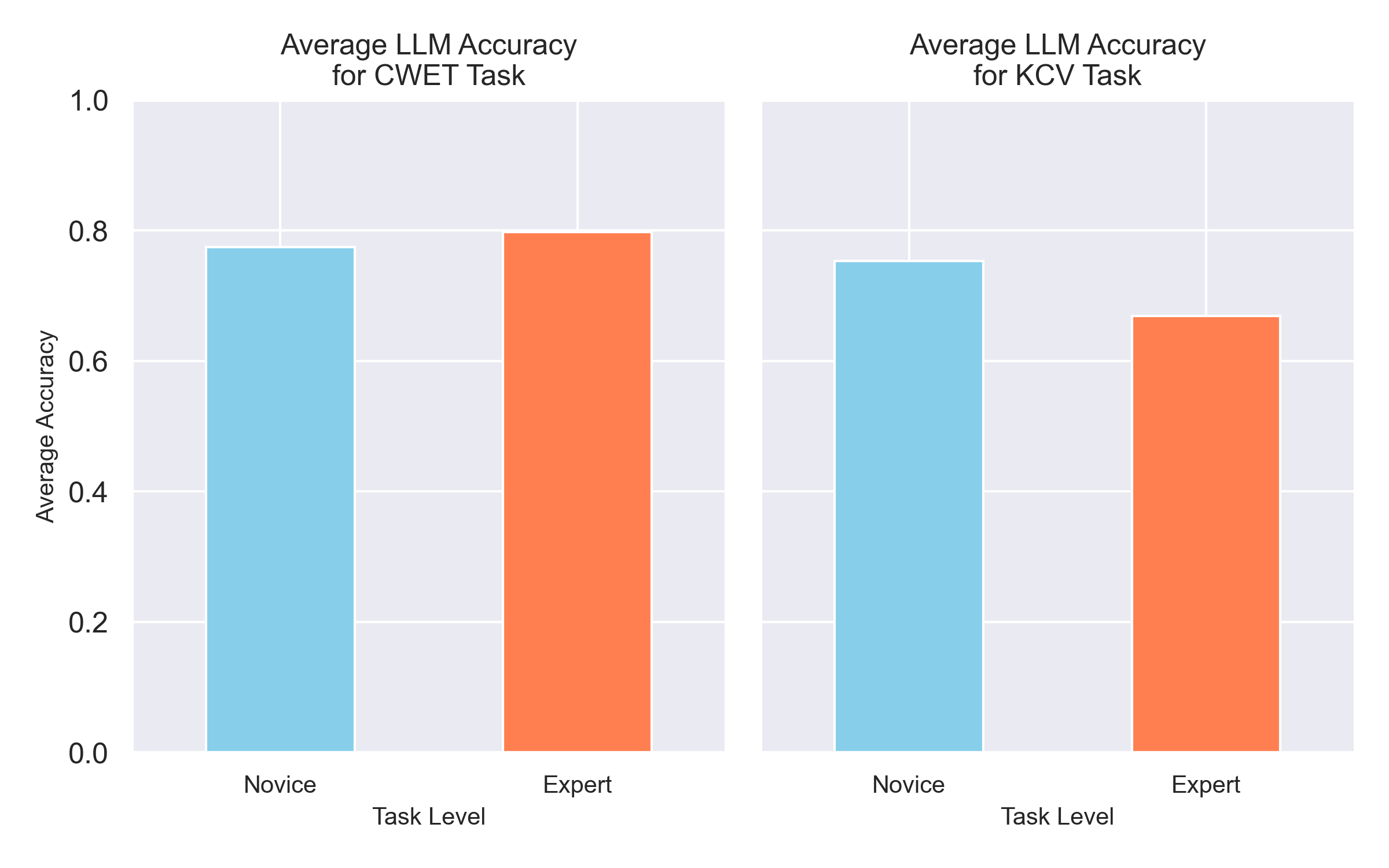} 
    \caption{Performance comparison of LLMs across two different levels of question difficulty.}
    \label{fig:task-level}
\end{figure}

During the question generation phase for the Multiple Choice Question (MCQ) and Boolean tasks, we developed distinct question sets reflecting two levels of expertise: expert (5+ years of experience in security) and novice (1-2 years of experience in security). The focus of the questions were on `fundamental concepts' for `novices' and `advanced technical details, problem-solving skills, and implications' for `experts'. This section analyzes the performance variations observed across these expertise levels.

Figure~\ref{fig:task-level} presents the average accuracy of all evaluated LLMs on the CWET and KCV tasks. Notably, LLMs demonstrate marginally superior performance on the ``expert'' questions for the CWET task, showing an improvement of 2.31\% compared to the ``novice'' security questions. This suggests that LLMs can leverage their extensive training data to excel even in specialized cybersecurity domains for tasks primarily reliant on existing knowledge and pattern recognition. However, as discussed in Section~\ref{sec:err-analysis}, they often fail to retrieve the correct source for more factual questions.

In contrast, the KCV task, which requires integrating new information and context-aware reasoning, reveals a different trend. Here, LLMs experience a significant decrease in accuracy, with a 5.44\% drop for ``expert'' tasks compared to ``novice'' tasks. This finding underscores a potential limitation of LLMs: while proficient at utilizing pre-existing knowledge, their capability to effectively incorporate and reason with novel information, particularly in complex and dynamic fields like cybersecurity, necessitates further investigation and refinement.

\subsection{Human Benchmark}
We conducted an experiment involving human subjects where participants were asked to answer a subset of the SECURE benchmark questions. 7 PhD students and 7 industry professionals working in cybersecurity and threat intelligence with years of experience ranging from 2-10 years were participants of this study.

We provided an MCQ form to each participant consisting of 75 questions from MAET, CWET, and KCV datasets divided into three sets, and were asked to provide their answers. We also collected their confidence level (high or low) when providing the answers and the total time taken for each set. Table \ref{tab:humaneval} shows the average performance of the human subjects on the dataset. Comparing the results with Table \ref{tab:results}, we can observe that large language models outperform human subjects on the three tasks. Most participants considered KCV a difficult task as it involves CVE document analysis before validating or refuting the given statement. On average, PhD students scored higher accuracy than industry professionals. While analysts scored an average accuracy of 72.2\% accuracy on all tasks, PhD students scored 77.3\% accuracy.

\begin{table}[]
\centering 
\caption{Average performance of human subjects on the subset of SECURE benchmark}
\label{tab:humaneval}
\resizebox{0.45\textwidth}{!}{%
\begin{tabular}{@{}lccc@{}}
\toprule
\textbf{Dataset} & \textbf{Accuracy (Average)} & \textbf{Confidence (Average)} & \textbf{Average time (mins)} \\ \midrule
MAET & 77\% & 76\% & 25 \\
CWET & 78\% & 81\% & 23 \\
KCV & 68\% & 72\% & 30 \\ \bottomrule
\end{tabular}%
}
\end{table}

\subsection{Recommendations}
Based on our findings from the analysis of LLMs on the SECURE benchmark, we provide the following targeted recommendations specifically designed to enhance ICS security:
\begin{enumerate}[leftmargin=*, noitemsep, topsep=0pt, partopsep=0pt]
\item \textbf{Confidence Calibration and Monitoring:} Implement robust mechanisms to monitor and adjust the confidence levels of LLMs when responding to sector-related queries. In ICS environments, inaccurate responses can have severe consequences. 

\item \textbf{Model Selection:} Unless open-source models demonstrate improved performance, prioritize the use of closed-source LLMs for problem-solving or out-of-distribution tasks (unless cost is a concern). These models have demonstrated superior performance in handling complex and unfamiliar scenarios.

\item \textbf{Encourage Detailed Explanations:} Require LLMs to provide detailed explanations or reasoning steps for their responses to security issues. This transparency not only improves trust but also allows human operators to understand the model's decision-making process. Implement this strategy where detailed insights are critical for accurate and informed decision-making.

\item \textbf{Addressing Hallucinations:} Integrate a human-in-the-loop process to review and validate LLM responses before implementation. This step is crucial to mitigate the risk of hallucinations, ensuring that the information and recommendations provided are reliable and actionable. This approach is especially important in environments where incorrect data can lead to significant operational disruptions or safety hazards.

\item \textbf{Improving Contextual Understanding:} Enhance LLMs' ability to interpret and respond accurately by improving their understanding of sector-specific terminology and scenarios. This can be achieved through better context framing and providing comprehensive background information relevant to the sector.
\end{enumerate}

\subsection{Ethical concerns}
All of the evaluation tasks in our proposed benchmark SECURE use publicly available threat information from credible sources like MITRE \cite{mitre-techniques} \cite{mitre-mitigations}, CVE \cite{cve2024}, CWE \cite{cwe1358}, and Cybersecurity and Infrastructure Security Agency (CISA) \cite{cisa_advisories}. None of the datasets contain any personal information, and they do not make sensitive judgments on social issues for bias, deception, or discrimination. 

\section{Customized LLMs}\label{sec:customizellm}

\subsection{Retrieval Augmented Generation} Retrieval Augmented Generation (RAG) is a method that combines the strengths of retrieval based models and generation based models \cite{lewis2020retrieval}. It consists of two major components: \textit{retriever and generator}. The retriever searches for relevant information from a large database or knowledge base, using  retrieval techniques to find the most relevant documents, paragraphs, or data points. Once relevant information is retrieved, the generator component (typically a language model like GPT) uses this information as context to generate a more accurate and contextually relevant response.

\begin{figure}[ht]
\centering
\includegraphics[width=0.90\linewidth]{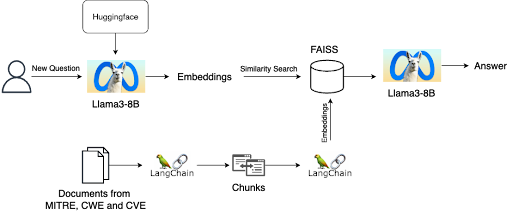}
    \caption{RAG Architecture}
    \label{fig:RAG}

\end{figure}

We use LLAMA3-8B as our generator model and design retriever to search for relevant information from sources like MITRE, CWE, and CVEs. We utilize the LangChain framework \cite{LangChain} for data-handling. The documents are segmented into chunks of 512 tokens with an overlap of 20 tokens using LangChain's RecursiveCharacterTextSplitter. This chunk size ensures that the model retains enough contextual information. The chunks are then transformed into vector embeddings using the "mixedbread-ai/mxbai-embed-large-v1" model \cite{emb2024mxbai}, and stored these embeddings in the FAISS database \cite{faiss}.

When a user queries the RAG model, it is first converted into vector embeddings using the same "mixedbread-ai/mxbai-embed-large-v1" model. These embeddings are then processed through LangChain's ConversationalRetrievalChain, which identifies the closest matches to the query from the stored documents. The matched documents, along with the query, are then fed into Llama3 to generate a coherent response. Figure \ref{fig:RAG} shows the block diagram of our RAG model.

\subsection{Fine-tuning}

Fine-tuning is the process of taking a pretrained model and training it further on a specific dataset to adapt it to particular tasks~\cite{zhang2024scaling}. The goal of this ‘tuning’ is to help the model learn the nuances of the new data. We follow the following steps:

\textbf{Data preparation:} We collect data using the web-graph data from Common Crawl \cite{commoncrawl}, and filter nodes of representative domains, based on their eigenvector centrality. We select the top 1 million nodes, ranked by highest centrality, for community detection, and scrap web pages from the identified domains within the community using Trafilatura \cite{barbaresi-2021-trafilatura}. We use this dataset for continual pre-training but exclude from the instruction fine-tuning stage.

We also utilize a a set of human-collected URLs to ensure the dataset is specific to Industrial Control Systems (ICS) by sorting and filtering the data extracted with Trafilatura. This step ensures the dataset focuses on ICS-specific content before proceeding to create the instruction dataset.

\textbf{Instruction generation and fine-tuning:} We apply the Self-Instruct method to generate instructions from the parsed text, using the default system prompt from the Self-Instruct paper \cite{wang2022self}. This step yielded 10,000 instructions. We generate responses using the Llama3.1 70B Instruct model, with 8-bit quantization, and a custom system prompt. We then apply instruction fine-tuning on the Llama3-8B Instruct model with 4-bit training.

\begin{table}[]
\centering 
\caption{Benchmark results on customized LLMs}
\label{tab:customizedLLMs}
\resizebox{0.35\textwidth}{!}{%
\begin{tabular}{@{}lccc@{}}
\toprule
\textbf{Model} & \textbf{MAET} & \textbf{CWET} & \textbf{KCV} \\ \midrule
Base-Llama3-8B & 82.1\% & 83.9\% & 82.8\% \\
RAG model & 86.6\% & 77.3\% & 60.5\% \\
Fine-tuned model & 84\% & 85\% & 76\% \\ \bottomrule
\end{tabular}%
}
\end{table}

\subsection{Results:} Table \ref{tab:customizedLLMs} shows the result of the benchmark. MAET, CWET, and KCV do not have source URLs provided as a context, and hence can accurately evaluate customized models performance against general LLMs. We can observe that both of the customized models perform better on MAET. Fine-tuned model outperform RAG on CWET. While fine-tuned model performs comparable to base model on KCV, RAG model has a significant drop. One key weakness of “Retrieve-Read” framework for RAG is that during the retrieval phase, the model can select irrelevant chunks and miss crucial information for accurate response. Hence, if the content is not supported by the retrieved context, the model faces hallucinations \cite{gao2023retrieval}. This could be one of the reasons behind lower performance in CWET and KCV. We will explore advanced RAGs framework in our future works.

\section{Limitations \& Future Work}\label{sec:limitations}

The reliance on standardized sources such as MITRE ATT\&CK, CWE, CVE, and CVSS introduces potential biases and coverage gaps in our benchmark datasets. While these sources are integral to the cybersecurity community and are frequently used in research, they are not entirely without limitations. In particular, issues such as incorrect versioning in CVE entries or incomplete coverage of attack techniques in MITRE ATT\&CK could impact the accuracy of SECURE's evaluation tasks.

To address this, we have validated the datasets through expert review and cross-referencing with additional sources. However, these measures cannot eliminate the inherent biases of the original databases. In future work, we will continue refining these datasets and exploring alternative data sources.

We focused on three distinct knowledge tasks (extraction, understanding, and reasoning) to assess the knowledge acquired by LLMs in the cybersecurity context, especially for ICS. While our current work involved designing six datasets centered on ICS security, we plan to extend this framework to encompass other areas of the cybersecurity industry. By adding more datasets, we aim to increase the evaluation ``breadth'' and provide a more comprehensive assessment. In future work, we also intend to explore various dimensions of trustworthiness, including toxicity, bias, adversarial robustness, out-of-distribution robustness, privacy, and fairness. These investigations are expected to reveal unidentified vulnerabilities and threats to the reliability of using LLMs in cybersecurity. 

\section{Conclusion}\label{sec:conclusion}
LLMs are rapidly transforming the landscape of artificial intelligence, offering vast potential applications in cybersecurity. However, concerns surrounding their reliability and understanding remain paramount, particularly within the sensitive and high-stakes cybersecurity domain. The SECURE benchmark, introduced in this paper, addresses these concerns by providing a comprehensive framework for assessing the capabilities of LLMs in a cybersecurity advisory context. Our benchmark encompasses diverse tasks, evaluating knowledge extraction and reasoning abilities to ensure a thorough assessment. Through our experiments and analysis, we demonstrate the potential of LLMs as valuable tools for cybersecurity professionals. However, our findings also highlight the need for caution, especially in handling complex reasoning tasks. By open-sourcing our benchmark datasets, we invite the broader research community to contribute to their refinement, thereby enhancing the reliability of LLMs and paving the way for their responsible and effective use in cybersecurity applications.

\newpage


\bibliographystyle{IEEEtran}
\bibliography{references}

\appendix

\subsection{Large Language Models}\label{appendix:models}

Large Language Models (LLMs) are pretrained autoregressive transformers \cite{vaswani2017attention} capable of generating one token at a time given a set of generated tokens as input. These are trained to maximize the probability of the next token given a history of tokens. We use the following LLMs in our evaluation:
\begin{enumerate}
    \item \textbf{ChatGPT models:} ChatGPT models by OpenAI have brought remarkable improvements in LLMs and sparked a trend for building more intelligent AI models. We use the following LLMs from OpenAI in our work, accessed via OpenAI’s API querying system \cite{gptapi}: 
    
   \textbf{ a)} \textbf{GPT-3.5} models can understand and generate natural language. We use the \textit{gpt-3.5-turbo} model that has been optimized for chat. It has a context window of 16385 tokens and the model was trained up to the cutoff time of September 2021.
    
  \textbf{  b)} \textbf{GPT-4} is a multimodal modal that can accept both text and image as inputs. This model excels in solving various linguistic tasks compared to GPT-3.5 model. We use the \textit{gpt-4-turbo} version which is optimized for chat. It has a context window of 128000 tokens. The model was trained on a dataset collected upto to December 2023.
    
   \textbf{ c)} \textbf{GPT-4o} is the latest model from openAI that can accept text and images as inputs. It is faster than GPT-4 Turbo but has the same context window size. We use this latest model to create the dataset.

\item \textbf{Llama3-70B and Llama3-8B:} Llama3 models \cite{llama370b} are a family of large language models developed by Meta, which is open-sourced to the researchers. These models are available in parameter sizes of 8B and 70B with both pre-trained and instruction-tuned (optimized for chat) available in public. These models input text only, and generate text and code only. We use these models through huggingface API. Llama3 models were released on April 18, 2024.

\item \textbf{Gemini \cite{gemini}:} Gemini models are a family of multimodal generative models released by Google which are capable of handling extremely long contexts, and take input as documents, video, audio and pictures. We use the Gemini-Pro 1.5 version of the LLM available through API.

\item \textbf{Mistral-7B and Mixtral 8x7B:} Mixtral 8x7b \cite{mixtral8x7b} is a mixture-of-experts model ( Mixtral-7B \cite{jiang2023mistral}), consisting of 8 distinct groups of parameters.  We use both of these models through huggingface API.

\end{enumerate}

\begin{table}[]
\caption{Comparison between different LLMs}
\label{tab:LLMCompare}
\resizebox{0.46\textwidth}{!}{%
\begin{tabular}{@{}lcclll@{}}
\toprule
\multicolumn{1}{c}{\textbf{Type}} & \textbf{Model} & \textbf{Provider} & \multicolumn{1}{c}{\textbf{Training data}} & \multicolumn{1}{c}{\textbf{\# of parameters}} & \multicolumn{1}{c}{\textbf{Context-window}} \\ \midrule
\multirow{3}{*}{Closed} & ChatGPT-3.5 & OpenaAI & Upto Sept. 2021 & - & 16385 \\
 & ChatGPT-4 & OpenaAI & Upto Dec. 2023 & - & 128000 \\
 & Gemini-Pro & Google & Upto Nov. 2023 & - & 128000 \\ \midrule 
\multirow{4}{*}{Open} & Llama3-70b & Meta & Upto Dec. 2023 & 70.6B & 8000 \\
 & Llama3-8b & Meta & Upto Mar. 2023 & 8B & 8000 \\
 & Mistral-7B & Mistral & Upto Sept. 2021 & 7B & 8192 \\
 & Mixtral-8x7B & Mistral & Upto Sept. 2021 & 46.7B & 32000 \\ \bottomrule 
\end{tabular}%
}
\end{table}

\vfill\null 

\subsection{Evaluation Metrics}\label{appendix:evalmetrics}

\begin{enumerate}[leftmargin=*, noitemsep, topsep=0pt, partopsep=0pt]
\item \textbf{MAET, CWET \& KCV:} For these three tasks, we utilize \textbf{accuracy} as the primary evaluation metric. Accuracy represents the percentage of questions answered correctly by the LLM.

\item \textbf{VOOD:} This out-of-distribution task presents models with True, False, and "Don't Know (X)" options. We consider a model's prediction accurate if it selects "Don't Know (X)," demonstrating its ability to recognize when it lacks sufficient information to answer.

\item \textbf{RERT:} We treat the RERT task as a summarization task, evaluating generated summaries against reference summaries using the \textbf{ROGUE-L} metric \cite{lin-2004-rouge}. ROGUE score is used in natural language processing (NLP) tasks to measure the similarity between a generated text and a reference text using overlapping units (such as n-grams, word sequences, or word pairs). In RERT, the LLMs are tasked with predicting the Risk Evaluation based on the provided vulnerability details that has a ground truth. We compute ROGUE-L by measuring the longest common subsequence (LCS) between the generated risk evaluation and reference ground truth. Given a ground truth summary $X$ of length $m$ and generated summary $Y$ of length $n$, to compute ROGUE-L, we first compute $R_{LCS} = \frac{LCS(X,Y)}{m}$ and $P_{LCS} = \frac{LCS(X,Y)}{n}$, where $LCS(X, Y)$ is the length of the longest common subsequence of X and Y. Then, ROGUE score can be computed as $ROGUE_L$ = $\frac{(1+\beta^2)R_{LCS}*P_{LCS}}{R_{LCS}+\beta^2*P_{LCS}}$, where, $\beta = \frac{P_{LCS}}{R_{LCS}}$.

\item \textbf{CPST:} The CPST task requires models to predict a real number within the range of 0 to 10. Given a ground truth of true CVSS scores, we compute its difference with the LLM predicted CVSS score and finally compute the average. This is measure of \textbf{mean absolute deviation (MAD)}, that quantifies the average distance between each data point in a set and the mean of that set. 
\end{enumerate}

\subsection{Prompt templates for evaluation}\label{appendix:prompt}

\subsubsection{Extraction Task}
\textbf{Example Prompt (Task: MAET)}

\begin{highlighted}{}
For the given question: Which protocol function can be disabled to prevent unauthorized device shutdowns?, and four options: A) DNP3 0x0D, B) HTTP GET, C) SMTP HELO, or D) FTP LIST, pick the best option as the answer, and return as either A, B, C or D. If you do not know the answer, return X. Choose the appropriate letter from A, B, C, D, or X as your answer. Please provide only the letter corresponding to your choice without any additional text or explanations.
\end{highlighted}{}

\textbf{Example Prompt (Task: CWET)}

\begin{highlighted}{}
For the given question: What is a common method used by attackers to bypass ATA password security?, and four options: A) Using a BIOS exploit, B) Hot swapping the drive, C) Encrypting the drive, or D) Using default passwords, pick the best option as the answer, and return as either A, B, C or D. If you do not know the answer, return X. Choose the appropriate letter from A, B, C, D, or X as your answer. Please provide only the letter corresponding to your choice without any additional text or explanations.
\end{highlighted}{}

\subsubsection{Understanding Task}
\textbf{Example Prompt (Task: KCV)}

\begin{highlighted}{}
You are given the following JSON data as context: $JSON\_FILE$. Based on the context, you have to analyze the following statement: PAN-OS versions 9.0.17 and later are unaffected by the vulnerability. and indicate whether the statement is True or False. Return your answer as either T (for True) or F (for False). If you do not know the answer, return X. Provide only the letter corresponding to your choice (T, F, or X) without any additional text or explanations.
\end{highlighted}{}

For this example we pass the $JSON\_file$ downloaded from this CVE-Link\footnote{\url{https://github.com/CVEProject/cvelistV5/blob/main/cves/2024/0xxx/CVE-2024-0008.json}}. 

\textbf{Example Prompt (Task: VOOD)}
In VOOD, we do not pass any context to the LLMs and inspect how the models perform on information that has not been seen by them during training.

\begin{highlighted}{}
For the given statement: The vulnerability described in CVE-2024-0011 allows for the execution of arbitrary code on the affected system., indicate whether the statement is True or False. Return your answer as either T (for True) or F (for False). If you do not know the answer, return X. Provide only the letter corresponding to your choice (T, F, or X) without any additional text or explanations.
\end{highlighted}{}

\subsubsection{Reasoning Task}
\textbf{Example Prompt (Task: RERT)}

\begin{highlighted}{}
You are given the following descriptions of a cyber incident Vulnerability Overview. Please perform an analysis and deduce the risk evaluation. Return your answer in one sentence in a format such as "Successful exploitation of this vulnerability could allow an attacker to inject arbitrary JavaScript into a user's web browser" for a single vulnerability, or "Successful exploitation of these vulnerabilities could cause a denial of service, disclosure of sensitive information, communication loss, and modification of settings or ladder logic" for multiple vulnerabilities.
3.2.1 Path Traversal CWE-22 There are multiple ways in LAquis SCADA for an attacker to access locations outside of their own directory. CVE-2024-5040 has been assigned to this vulnerability. A CVSS v3.1 base score of 7.8 has been calculated; the CVSS vector string is (AV:L/AC:L/PR:N/UI:R/S:U/C:H/I:H/A:H). A CVSS v4 score has also been calculated for CVE-2024-5040. A base score of 8.5 has been calculated; the CVSS vector string is (CVSS4.0/AV:L/AC:L/AT:N/PR:N/ \\ UI:P/VC:H/VI:H/VA:H/SC:N/SI:N/SA:N).
\end{highlighted}{}

\textbf{Example Prompt (Task: CPST)}
\begin{highlighted}{}
Calculate the CVSS v3 score for the following CVSS v3 Vector String: \\ AV:L/AC:L/PR:N/UI:R/S:U/C:H/I:H/A:H. Return your answer as a single number between 0.0 and 10.0. Provide only the number corresponding to your answer without any additional text or explanations.
\end{highlighted}{}

\vfill\null 

\subsection{LLM Reasoning Example}
Example Llama3-70b reasoning for one question in the KCV task
\begin{highlighted}{}
To determine the accuracy of the statement, I analyzed the provided JSON data, which describes a CVE record for CVE-2024-0011. The vulnerability is a reflected cross-site scripting (XSS) issue in the Captive Portal feature of Palo Alto Networks PAN-OS software.

The description of the vulnerability states: "A reflected cross-site scripting (XSS) vulnerability in the Captive Portal feature of Palo Alto Networks PAN-OS software enables execution of malicious JavaScript (in the context of an authenticated Captive Portal user’s browser) if a user clicks on a malicious link, allowing phishing attacks that could lead to credential theft."

The key points to note are:

1. The vulnerability is a reflected XSS issue, which means it allows an attacker to inject malicious JavaScript code into a user's browser.
2. The vulnerability is limited to the context of an authenticated Captive Portal user's browser.
3. The goal of the attack is to perform phishing attacks that could lead to credential theft.

Nowhere in the description or the provided data is there any indication that the vulnerability allows for the execution of arbitrary code on the affected system. The vulnerability is primarily a client-side issue, affecting the user's browser, rather than a server-side issue that could lead to arbitrary code execution on the system.

Therefore, based on the analysis of the provided data, I conclude that the statement is False.    
\end{highlighted}

\end{document}